# Physical variability in meter-scale laboratory CO$_2$ injections in faulted geometries

*Malin Haugen[1], Lluís Saló-Salgado[2,3], Kristoffer Eikehaug[1], Benyamine Benali[1], Jakub W. Both[4], Erlend Storvik[4], Olav Folkvord[1], Ruben Juanes[2,3], Jan Martin Nordbotten,[4,5] and Martin A. Fernø[1,5]*

1 Department of Physics and Technology, University of Bergen
2 Department of Civil and Environmental Engineering, Massachusetts Institute of Technology, Cambridge, 02139, MA, USA
3 Earth Resources Laboratory, Department of Earth, Atmospheric and Planetary Sciences, Massachusetts Institute of Technology, Cambridge, 02139, MA, USA
4 Center for Modeling of Coupled Subsurface Dynamics, Department of Mathematics, University of Bergen
5 Center for Sustainable Subsurface Resources, Norwegian Research Center, Postboks 22 Nygårdstangen, 5838 Bergen

**Abstract**

Carbon, capture, and storage (CCS) is an important bridging technology to combat climate change in the transition towards net-zero. The FluidFlower concept has been developed to visualize and study CO$_2$ flow and storage mechanisms in sedimentary systems in a laboratory setting. Meter-scale multiphase flow in two geological geometries, including normal faults with and without smearing, is studied. The experimental protocols developed to provide key input parameters for numerical simulations are detailed, including an evaluation of operational parameters for the FluidFlower benchmark study. Variability in CO$_2$ migration patterns for two different geometries is quantified, both between 16 repeated laboratory runs and between history-matched models and a CO$_2$ injection experiment. The predicative capability of a history-matched model is then evaluated in a different geological setting.

**Keywords:** Geologic carbon sequestration; CO$_2$ storage; FluidFlower; laboratory; numerical simulations

**Introduction**

As the prices of renewable energy are decreasing, carbon, capture, utilization, and storage represent a bridging technology to combat climate change in the transition period towards net zero. Geological carbon sequestration (GCS) could contribute to the energy transition by tackling emissions from existing energy assets, providing solutions in some of the sectors where emissions are hardest to reduce (like cement production), supporting the rapid scaling up of low-emissions hydrogen production, and enabling some CO$_2$ to be removed from the atmosphere through bio energy and direct air capture with GCS (IEA, 2021).

Carbon sequestration is based on the principle that the injected CO$_2$ become less mobile over time by porous media trapping mechanisms, where the relative importance between the governing processes depends on the subsurface conditions. In-situ visualization of CO$_2$ injection and the trapping mechanisms are valuable for understanding the fluid flow and migration patterns during geological CO$_2$ sequestration (GCS), and the authors believe it is also important for enhancing public understanding and acceptance about CO$_2$ storage and security. The laboratory experiments presented in this paper are relevant for geological carbon storage as the main mechanisms are sustained, including capillarity, dissolution and convective mixing, and it represent a unique possibility to test our simulation skills, because in contrast to the subsurface, here we can compare predictions to observations.

The FluidFlower concept links GCS research and dissemination through experimental rigs constructed at University of Bergen (UoB) that enable meter-scale, multiphase, quasi two-dimensional flow on complex, yet representative geological geometries. Different geological geometries are constructed using unconsolidated sands, and simulation models based on the experimentally studied geometry provide key input parameters for numerical simulations. Gaseous and aqueous forms of CO$_2$ are distinguished with pH-sensitive dyes, and the multiphase fluid flow development is captured with time-lapse imaging. The in-house developed software DarSIA (Darcy Scale Image Analysis toolbox, Nordbotten et al., this issue) is utilized to analyze the images to quantify key parameters and variability in the experimentally observed CO$_2$ migration patterns.

The goal of this study is three-fold: First, provide and discuss petrophysical properties, including single- and multiphase flow parameters, to enable the evaluation of how accessibility of such data impacts numerical simulations of geologic CO$_2$ storage (Saló-Salgado et al., this issue). Second, evaluate operational parameters and processes for the FluidFlower benchmark study, as specified in Nordbotten et al. (2022). Third, perform dedicated laboratory experiments for history matching (Saló-Salgado et al., this issue), and determine the predicative capability of the validated and physics-based model in a different geological setting.

The paper is divided into three sections: 1) analysis and measurements of physical properties of the unconsolidated sand used to build the studied geometries, 2) an overview of the fast-prototyping approach to evaluate impact of key operational parameters, with a discussion on physical variability between repeated $CO_2$ injections in two geological geometries, and 3) numerical modeling of experimental $CO_2$ injections with history matching and simulation with comparison to physical experiments.

## 1 Physical properties of the quartz sand - experimental protocol and measured values

When sediments (particles) accumulate they form sedimentary deposits which compose layers of rock. Within a deposit, the individual particles vary in size, shape, etc., and hence the layer they constitute has possess certain macroscopic properties such as porosity and permeability (Krumbein and Monk, 1942). These mass properties will vary with, among others, the combination of properties of the particles and the conditions of packing (Krumbein and Monk, 1942). The geologically important characteristics of sediments might be described using six measurable quantities: size, shape (sphericity), roundness (angularity), mineral composition, surface texture and orientation. The following sections describe how physical properties of the unconsolidated sand were measured together with the result. Some of these properties are further used as input to simulation models for performing history match of $CO_2$ injection in the Medium FluidFlower experiments presented here (Saló-Salgado et al., this issue), and also in the benchmark study specified in Nordbotten et al. (2022) and reported in Flemisch et al. (this issue) and Fernø et al. (this issue).

### 1.1 Preparation of sand

The studied geological geometries were constructed using unconsolidated quartz sand from a commercial Norwegian supplier. Based on the stated and available grain size ranges, desired grain size ranges for this study were chosen according to the Wentworth scale (Wentworth, C.K., 1922) (**Table 1**). The received sands were sieved prior to building the geological geometries to achieve increased control of the used grain size ranges. Dry wire-mech sieves (from Glenammer) staked on a mechanical shaker were used, before the sands were washed in a two-step process: 1) Rinsed with tap water to remove fine material, 2) Acid washed using HCl (**Table 4**) until carbonate impurities were dissolved, and no more $CO_2$ bubbles were observed (varying from 24 to 72 hours for the different sand types). The acid was then neutralized, and the sand was rinsed with tap water. After washing, the sands were dried at 60ºC for at least 24 hours and stored in clean 15 L plastic containers until use. Sands named ESF, C, E and F were used to build the geometries investigated here, whereas all sand-types (C-G + ESF) were used in FluidFlower benchmark geometry described elsewhere (Fernø et al., this issue).

*Table 1: Desired grain-size range of sand used in this study and size of wire-mesh sieves used*

| Sand ID | Desired grain size range [a] [mm] | Desired grain size range [a] [phi] | Grade [a] |
|---|---|---|---|
| **ESF** | 0.13 to 0.36 [b] | 2.9 to 1.5 | Fine sand |
| **C** | 0.5 to 0.71 | 1.0 to 0.5 | Coarse sand (lower) |
| **D** | 0.71 to 1.0 | 0.5 to 0.0 | Coarse sand (upper) |
| **E** | 1.0 to 1.41 | 0 to -0.5 | Very coarse sand (lower) |
| **F** | 1.41 to 2.0 | -0.5 to -1.0 | Very coarse sand (upper) |
| **G** | 2.0 to 2.8 | -1.0 to -1.5 | Granule gravel |

[a] Sieved range and grade according to Wentworth class (Wentworth, C.K., 1922)
[b] Size range provided by supplier. ESF sand was not sieved, only washed and dried to maintain fine particles

### 1.2 Textural properties of the sand

Key sand grain properties were derived from segmented, binary microscopic images (Zeiss, Axio Zoom.V16) to obtain the distributions of sand grain width, length, and sphericity using Python and OpenCV functions (see **Table 2**). The most important textural properties of natural clastic sediments can be expressed as four quantities: 1) grain size, 2) sorting, 3) sphericity, 4) roundness (angularity), and 5) packing (Beard and Weyl, 1973). The grain size and sorting are measurable, and important factors in the porosity and permeability, however packing (grain arrangement) of unconsolidated sand is difficult to measure and to assess its impact on porosity and permeability (Beard and Weyl, 1973).

The grain size (mm) was converted to so-called phi scale, where (*grain size*) $_{in\ phi}$ = $-log_2$ * (*grain size*) $_{in\ mm}$ (Krumbein, 1936). The grain size measurements and sphericity (**Table 2**) demonstrate that many grains were noncircular and only pass through the sieves when oriented vertically. The grain size width, rather than grain size, is therefore used as in the following discussion. Compared to the desired grain-size range (**Table 1**), each sand type had a larger distribution (**SI.fig. 1**) than expected from the sieving process. According to Folk and Ward (1957), standard deviation (std) of grain size in phi-scale is a measure of sorting, and according to their classification, std of 0.71 – 1.00 (Sand ESF) is moderately sorted, while 0.35 – 0.50 (Sand G) is well sorted and <0.35 (Sand C, D, E and F) is very well sorted.

*Table 2: Calculated properties of samples of the sieved sand, including width, length, and sphericity.*

| Sand ID | # Grains analyzed | Grain size width | | Grain length | | Sphericity [a] | |
|---|---|---|---|---|---|---|---|
| | | geomean ± std [mm] | mean phi-scale ± std | geomean ± std [mm] | mean phi-scale | geomean | std |
| **ESF** [b] | 100 | 0.16 ± 0.12 | 2.62 ± 0.82 | 0.26 ± 0.17 | 1.94 | - | - |
| **C** | 1127 | 0.67 ± 0.09 | 0.58 ± 0.18 | 0.86 ± 0.17 | 0.21 | 0.69 | 0.10 |
| **D** | 1190 | 1.05 ± 0.14 | -0.07 ± 0.19 | 1.36 ± 0.23 | -0.45 | 0.68 | 0.11 |
| **E** | 1000 | 1.44 ± 0.17 | -0.53 ± 0.17 | 1.87 ± 0.31 | -0.90 | 0.64 | 0.14 |
| **F** | 1112 | 1.78 ± 0.26 | -0.83 ± 0.22 | 2.25 ± 0.43 | -1.17 | 0.59 | 0.18 |
| **G** | 959 | 2.56 ± 0.56 | -1.35 ± 0.38 | 3.27 ± 0.70 | -1.71 | 0.39 | 0.26 |

[a] Two-dimensional sphericity where circular = 1.0. Calculated as $(4 * \pi * Area)/(Circumference^2)$, where area and circumference are calculated using the OpenCV functions contourArea and arclength, respectively.
[b] ESF grains were measured manually using a distance functionality in the Zeiss software because the grains were not possible to separate due to electrostatic forces.

## 1.3 Experimental setups and procedure

The petrophysical properties porosity ($\varphi$), absolute permeability (K) and endpoint relative permeability (kr) of the unconsolidated sand were measured (**Table 3**) using a sand column (**Fig. 1**). Custom-made end pieces (polyoxymethylene) with a Y- shaped passage enabled absolute pressure measurement (ESI, GSD4200-USB, -1 to 2.5 barg) at each end face, with milling for a round metal mesh and paper filter against the sand to maintain sand column integrity. A differential pressure transducer (Aplisens PRE-28, 0-2.5 bar) was positioned 154mm from the top and bottom of the tube and measured the differential pressure across a 171 mm section of the sand column. The procedural steps for the petrophysical measurements are detailed below, and different fluids were injected through the sand column throughout the procedure. Note that all bottles and outlets were at the same height to ensure a stable system with no flow in or out of the sand column when fluid injection was stopped. Tubes were fixed in place to mitigate influence on measured sand column weight and produced fluids during endpoint relative permeability measurements.

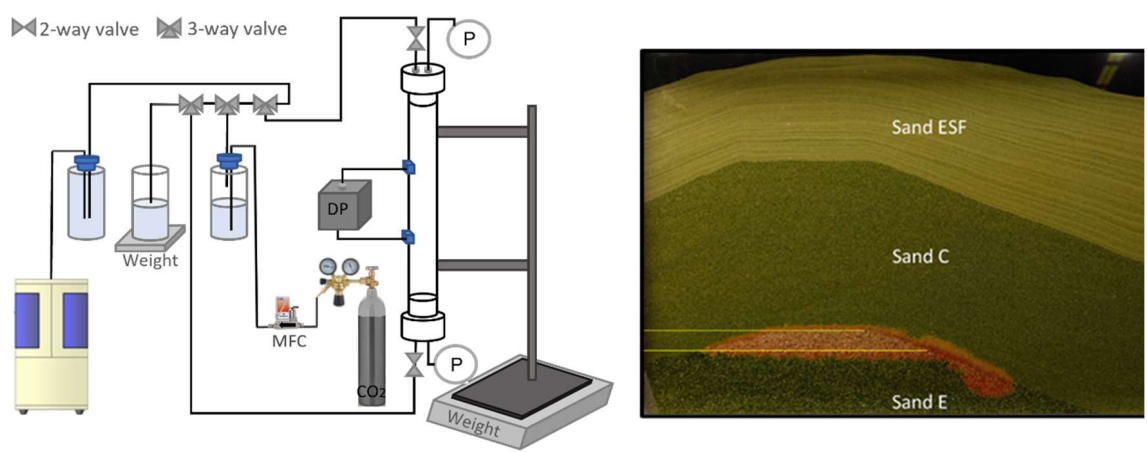

*Fig. 1: Left: schematic of the experimental set-up used for porosity, absolute permeability, and relative permeability. A vertical Plexiglas tube (length 478.0 mm and diameter 25.8 mm) with metal mesh and 0.2 µm paper filter against the sand maintained sand column integrity. ESI pressure transducers (-1 – 2.5 barg) monitored inlet and outlet pressure and temperature. An Aplisens differential pressure transducer (0 – 2.5 barg) recorded the differential pressure in the middle of the sand column. Quizix QX pump controlled the volumetric rate of injected aqueous phases, whereas a Bronkhorst El-Flow Prestige mass flow controller with a maximum rate of 10ml/min was used for the $CO_2$ injection. Right: Image with example of geometry in the Medium FluidFlower (**Fig 2**) used for gas column break-through experiments to obtain capillary entry pressure (height of gas column, distance between horizontal yellow lines), here for sand C.*

*Porosity (φ) measurement*

Porosity was calculated for all sand types (**Table 3**) using the following procedure:

1. The vertical tube, with bottom end-piece and filter attached, was filled with degassed, deionized (DI) water to a predetermined and known water height.
2. Dry sand was poured into the water-filled tube and settled at the bottom. Approximately 20 mm sand was added each time, and the tube was gently tapped during sand settling. Sand was added until the sand column height reached the predetermined level.
3. Water was constantly removed from the top of the tube during sand settling and the volume of the extracted water was measured cumulatively.
4. Prior to attaching the top end-piece, the tube was filled with DI (to minimize air between the end-piece and the sand) and the filter placed on top of the sand. Tubing attached to the end-piece was loosened to enable the displaced water to escape as the end-piece was placed into the tube without asserting force to the sand.

The extracted water volume in step 3. equals the sand grain volume, and the porosity was calculated from the ratio of the pore volume (bulk volume – grain volume) to bulk volume.

*Absolute permeability (K) measurement*

Absolute permeability (**Table 3**) was measured subsequent to the porosity measurement described above with the following procedure:

5. Degassed DI water was injected from the bottom with a high volumetric injection rate (650 ml/h for 3 hours) to remove trapped gas bubbles, if any.
6. Flow direction was reversed (top endpiece used as inlet) and degassed DI water was injected using 5 ascending and descending constant volumetric injection rates between 200 – 600 ml/h, with 100 ml/h increments for 10 min (stable differential pressures were achieved).
7. Differential pressures versus injection rate were recorded and used for calculation of absolute permeability using Darcy's law.

*Unsteady-state endpoint relative permeability ($k_r$) measurements*

The unsteady-state end-point relative permeabilities (**Table 3**) during both drainage and imbibition used $CO_2$-saturated water and $CO_2$. The average sand column end-point fluid saturation was calculated from volumetric and weight measurements. The following procedure was followed:

8. After absolute permeability measurements (step 5-7 above), degassed DI water was miscibly displaced with $CO_2$-saturated DI water. This step was performed to avoid dissolution of gaseous $CO_2$ gas in the water phase during drainage $CO_2$ injection.
9. End-point drainage: $CO_2$ gas was injected from the top with a constant rate of 10 $ml_s$/min (10 ml/min @ standard conditions: 20 deg C + 1013 millibar). The absolute inlet and outlet pressure (sands E, F, G) or differential pressure (sands ESF, C, D), were recorded during drainage. Water production was monitored during drainage (by weight), and $CO_2$ was injected until no further water production was recorded. The weight of the partially water saturated sand column was recorded. Endpoint relative permeability to $CO_2$ was calculated based on the pressure differential between 10 ml/min and 0 ml/min.
10. End-point imbibition: After the drainage process was completed, the injection was switched back to $CO_2$ saturated water and ramped up to a constant volumetric rate of 600ml/h, also injected from the top. The pressures and weight of sand column were recorded during injection. When no additional gas was produced for 60 min, the final sand column weight was recorded and an injection cycle to measure water endpoint permeability was conducted.

*Capillary entry pressure measurement*

The capillary entry pressure to gas (**Table 3**) was experimentally measured for each sand type based on observed gas column break-through experiments in the Medium FluidFlower (**Fig. 2**). Different geometries were investigated, including anticlines with an "inverse V shape" top to an "inverse U shape" top (cf. **Fig. 1**), constructed using the sedimentary protocol detailed in section 2.3. The porous media was saturated with pH-sensitive dye nr.1 (**Table 4**), and gaseous $CO_2$ was injected at 10 $ml_s$/min min (10 ml/min @ standard conditions: 20 deg °C and 1013 millibar). The gas accumulation and column height were monitored with time-lapsed imaging to detect the maximum gas column height between the different sand layers and under the sealing layer (ESF). The observed gas column height (in meter) was converted to pressure to provide the entry pressure for each sand type.

## 1.4 Petrophysical properties

Measurements of the petrophysical properties were performed as detailed in section 1.3 and the results are presented in **Table 3**.

*Table 3: Calculated porosity, absolute permeability, endpoint relative permeability for gas and water and capillary entry pressure for each sand*

| Sand ID | Porosity [a] | K[D] [b] | Endpoint gas ($CO_2$) | | Endpoint water | | Capillary entry pressure |
|---|---|---|---|---|---|---|---|
| | | | $S_{wi}$ | $k_{rel.gas}$ | $1-S_g$ | $k_{rel.water}$ | Pc [mbar] |
| **ESF** | 0.44 | 44 | 0.32 | 0.09 | 0.86 | 0.71 | 15.0 |
| **C** | 0.44 | 473 | 0.14 | 0.05 | 0.90 | 0.93 | 3.3 |
| **D** | 0.44 | 1110 | 0.12 | 0.02 | 0.92 | 0.95 | 0.9 |
| **E** | 0.45 | 2005 | 0.12 | 0.10 | 0.94 | 0.93 | 0.26 [c] |
| **F** | 0.44 | 4259 | 0.12 | 0.11 | 0.87 | 0.72 | 0.10 [c] |
| **G** | 0.45 | 9580 | 0.10 | 0.16 | 0.94 | 0.75 | 0.01 [c] |

[a] average based on constructing each sand pack twice with the same sand sample
[b] average calculated from one ascending and descending cycle (9 datapoints) for each sand
[c] calculated from the power function of the trendline in the log-log plot of Pc against geometric mean grain width for ESF, C and D sand because gas accumulation was not detectable during the capillary entry pressure measurements.

Comparing grain size versus sorting (Beard and Weyl, 1973) for sand D, C and ESF, the measured porosities (**Table 3**) are in the range of dry-loose and not wet-packed, however, for dry-loose packed sand they observed increasing porosity with decreasing grain size, while for wet-packed sand, the porosity remained about the same regardless of grain size, and this is also what was observed in the measured porosity presented here (**Table 3**). The permeability for each sand type of unconsolidated sand varies with the square of an average diameter (Krumbein and Monk (1942), and references therein), and this is also the case for the measured values presented here (**Table 3**). Dataset from the literature is compared to the measured values in **SI.fig. 2**. The capillary entry pressure are sensitive to orientation and packing of grains, and we observe differences between horizontal layers and vertical features (like a fault zone) in a geometry (see **SI.fig. 3**). The sensitivity is used in the history matching of capillary entry pressures in the numerical modeling of the experimental $CO_2$ injections.

The inner diameter of the cylindrical tube used during permeability measurements should be minimum 8 - 10 times the maximum particle size of the tested sand column (Chapuis, R.P., 2012). The geometric mean width and length for Sand-F and Sand-G (**Table 2**) is on this threshold value (8-10 times) with an inner tube diameter of 25.8 mm. Hence, this adds to the uncertainty to the measured permeability values as the relatively large grain size versus tube diameter may lead to poor packing conditions and preferential flow along tube walls (Chapuis, R.P., 2012).

## 2 Experimental $CO_2$ injection

The meter-scale Medium FluidFlower rig represents a physical asset to evaluate the parameter and operational space for the development of room-scale FluidFlower experiments and the International FluidFlower Benchmark initiative (Fernø et al., this issue, Eikehaug et al., this issue, Nordbotten et al. 2022). The rigs were designed and built at the University of Bergen to facilitate a fast-prototyping approach.

### 2.1 Experimental set-up and materials

The design of the Medium FluidFlower (**Fig. 2**) enables construction of a variety of geological geometries (using unconsolidated sands) and repeated injection scenarios to evaluate reproducibility between identical experiments without removing the sand and rebuilding the geometry. The porous media was constructed using unconsolidated sand and held in place by two optically transparent panels in the front and the back, glued together to an aluminum frame with spacing of 10 mm. The flow cell has no-flow boundaries in the bottom and sides, whereas the top is open and in contact with atmospheric pressure. The size and design of the medium rig allow relatively rapid testing of key operational conditions, e.g., different pH-indicator mixes, injection protocols, constructing different geological structures and the effect of degassed aqueous phase. Technical and mechanical properties of the FluidFlower rigs are detailed in Eikehaug et al., (this issue).

Two different geometries (termed 'Albus' and 'Bilbo') were studied in the Medium FluidFlower rig. In addition to the rig, the experimental set-up (**Fig. 2**) consists of throttle valve and Swagelok valves (2- and 3-way), mass flow controller (MFC) (EL-FLOW Prestige FG-201CV 0-10 mls/min, BronkHorst) calibrated for $CO_2$, pressurized $CO_2$ canister including pressure regulator, gas trap, ColorChecker (X-Rite), and camera with time-lapse function (Albus geometry: Sony ZV-1, and Bilbo geometry: Sony A7III, lens SAMYANG AF 45 mm F1.8). The high-resolution images (5472x3080 and 7952x4472 pixels for Albus and Bilbo geometries respectively) enable monitoring and analysis of multiphase flow dynamics with single grain identification and is one of the main measurements in this set-up.

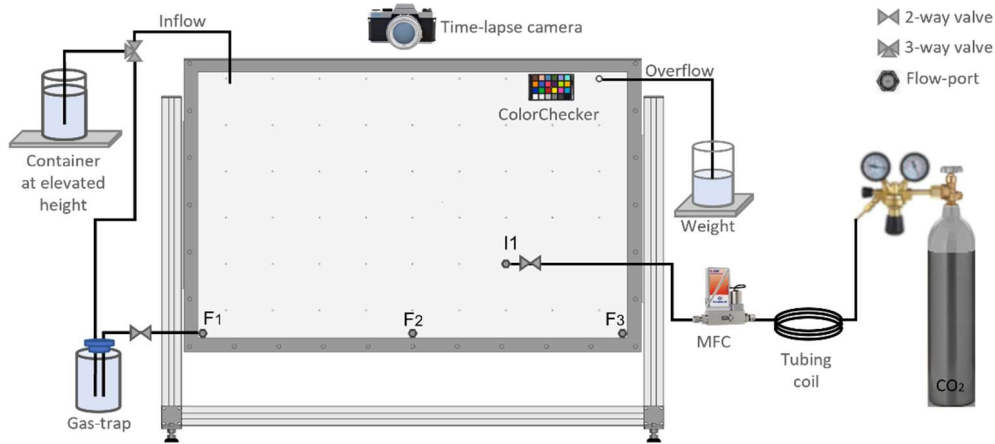

*Fig. 2: Schematic overview of the Medium FluidFlower and the equipment. Fluids are injected through gravity (container at elevated height) or using a pump. During flushing and resetting the flow-cell, the flush-ports (F1 – F3) are utilized, and as a preventive measure to avoid unwanted air in the rig, a gas-trap was included. For gas injection, the set-up consists of a $CO_2$ calibrated mas flow controller (MFC), $CO_2$ tank with regulator and a coil of tubing mitigate pressure fluctuations caused by the gas regulator. Inflow (tubing end) and outflow (always-open port) provide a constant hydraulic head and provide information about volume displaced if logged. A camera with time lapse function is used to acquire high resolution images of the dynamics, and the ColorChecker ensure correction of colors during image analysis to mitigate changes in illumination. All flow ports are 1/8-inch NPT to 1/8 Swagelock, with inner diameter of 1.8 mm.*

*Fluids*

Fluid properties is summarized below (**Table 4**) and detailed elsewhere (Eikehaug et al., this issue). The porous media is initially fully saturated an aqueous pH-indicator mix, referred to as "formation water". The $CO_2$ is injected as dry gas and will partially partition into the formation water to form "$CO_2$ saturated water". The image analysis (section 2.5) can distinguish between the different phases from the added pH sensitive dyes (gas and aqueous phases are originally transparent). Three different pH-indicator mixes (**Fig. 3**) are used to test visual impact on tracking $CO_2$ gas migration (as absence of color) and $CO_2$ saturated water, and further how this is influenced by the image segmentation using DarSIA (Nordbotten, et al., this issue).

*Table 4: Fluids and pH-indicator mixes with their expected pH values.*

| Fluid (phase) | Composition | pH | Usage |
|---|---|---|---|
| Acid (aq) | Tap water with<br>- Above 0.1 M hydrochloric acid (HCl) | < 2 * | Remove carbonate impurities during sand cleaning |
| pH-indicator mix 1 (aq) | Deionized water with<br>- 0.14 mM bromothymol blue ($BTB^-$)<br>- 0.43 mM methyl red ($MRe^-$)<br>- 0.10 mM hydroxide ($OH^-$)<br>- 0.67 mM sodium ions ($Na^+$) | ~ 8.3 | Initial formation fluid to enable detecting $CO_2$ dissolution in the aqueous phase |
| pH-indicator mix 2 (aq) | Deionized water with<br>- 0.14 mM bromothymol blue ($BTB^-$)<br>- 0.10 mM hydroxide ($OH^-$)<br>- 0.24 mM sodium ions ($Na^+$) | ~ 8.3 | Initial formation fluid to enable detecting $CO_2$ dissolution in the aqueous phase |
| pH-indicator mix 3 (aq) | Deionized water with<br>- 0.75 mM bromothymol blue ($BTB^-$)<br>- 1.00 mM hydroxide ($OH^-$)<br>- 0.67 mM sodium ions ($Na^+$) | < 10.4** | Initial formation fluid to enable detecting $CO_2$ dissolution in the aqueous phase |
| $CO_2$ (g) | 99.999% - 5.0 purity | - | Injected as gaseous phase |
| Lye solution (aq) | Deionized water with<br>0.48 mM sodium hydroxide (NaOH) | < 10.7** | Reset/clean the system after experiments (flush out $CO_2$ saturated water and trapped free gas) |

* Maintained below 2 until stable. ** Depending on exposure to air

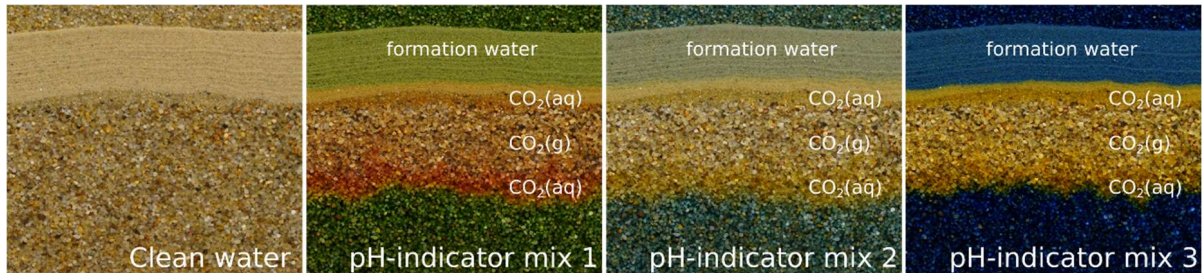

*Fig. 3: Illustration of different pH-indicator mixes used in the experiments, and their colored response to gaseous and dissolved $CO_2$ as supposed to the formation water. From left to right: Image of a layered sand formation saturated with clean water; same formation but saturated with pH-indicator mixes 1-3 in the presence of $CO_2$.*

The pH of the formation water impacts the equilibrium between gaseous $CO_2$ and the aqueous phase. It is therefore expected that experiments with pH-indicator mix 3 (pH approximately 10) will have increased $CO_2$ dissolution rate (and less mobile $CO_2$) compared with mix 1 and 2 (pH approximately 8), resulting in less lateral spreading of the $CO_2$ saturated water. The increased $CO_2$ concentration in $CO_2$ saturated water also increase its density and influence gravitational-dominated flow processes.

## 2.2   Geological models

The geological models used in this study (**Fig. 4**) were motivated by relationship between fluid flow and the presence of folds and faults in sedimentary rocks and basins. Two layered geometries with folds and faults with different properties (detailed below) are included to test our ability to incorporate and investigate such geological features using unconsolidated sand, following three guiding principles (Fernø et al., this issue): 1) enable realistic $CO_2$ flow pattern and trapping scenarios with increasing modeling complexity, 2) being sufficiently idealized so that the sand facies can be reproduced numerically with high accuracy, and 3) be able to operate, monitor and reset the fluids within a reasonable time frame.

Both geometries include two main reservoir sections separated by a lower seal (ESF sand) unit and are overlayed by a regional sealing unit at the top of the geometry. For geometry A (**Fig. 4**, top), the layers are folded and have two anticlines, one towards the left edge of the rig and one to the right. The right anticline is offset by a listric normal fault, causing a discontinuity in the lower sealing unit. The fault is represented by a fault plane (~28 mm, sand C) with an 80 ° dip angle and lower permeability compared with the main reservoir (sand F). Features of the listric normal fault cause a rollover anticline on the footwall (right side of the fault), representing another structural trap. The upper reservoir section consists of layers of high permeability sand (*Upper F and E, and Middle F*) intersected by a layer of lower permeability sand (*Upper C*), and the lower reservoir section consists of uniform high permeability sand (*Lower F*). For geometry B (**Fig 4**, bottom), one can imagine that there has been folding that created an anticline across the field of view, which later has been faulted. The structure has an erosional surface overlayed by a *Top Regional Sealing* unit (ESF sand). Note that the fault zone is not included, and the dip (70-80 degrees) and throw changes downwards in the reservoir layers. In contrast to geometry A, the *Lower Sealing* layer between the two reservoir sections in geometry B is continuous, and the sealing properties are therefore expected to be maintained. This could be described as smearing or drag folding and is a common feature when the clay content is high and is a feature which could have a large impact on fault transmissibility and fluid flow in the reservoir. Note that the drag/smear feature does not contain clay in geometry B, and because of increased leakage potential along the vertical no-flow boundaries, the layers dip downwards towards the edges to reduce this risk.

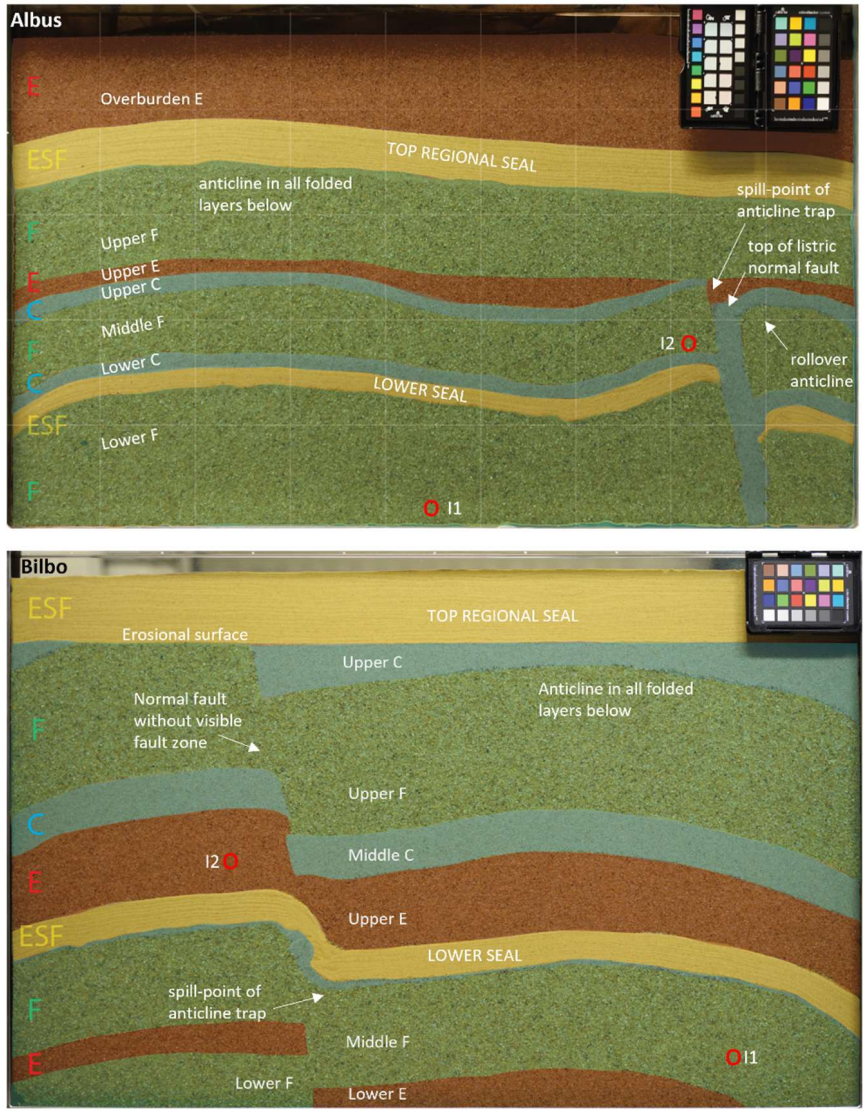

*Fig. 4: The Albus geological model (top) and the Bilbo geological model (bottom) Colors represent sand properties according to **Table 2**, and the geometries are overlayed by a fluid height of a few cm. Ports used for $CO_2$ injection are marked with red circles. Albus geological model is overlain with a 1000 x 1000 mm cartesian grid with [0,0] in the lower left corner.*

## 2.3   Procedure for constructing geometries

The size and operational capabilities of the Medium FluidFlower makes it a valuable asset to test and develop procedures and tools for constructing geometric features. A detailed sketch of the desired geological geometry is drawn on the front panel (**Fig. 5**), and dry sand with predetermined grain size is poured from the top into the water-filled flow cell. Hence, the geometry is built from bottom to top, and excess water is produced through the "overflow" port (**Fig. 2**) to maintain a constant water level during construction. Increased complexity may be achieved by 1) increasing the number of sand types and creating sequences with many layers (i.e., each layer consists of one sand type), or 2) adding features such as faults or heterogeneities with different properties.

In this work, three tools were designed to construct a fault using unconsolidated sand:
- 'Angle tools' for sharp edges/faults (used in **Fig. 5**). Made from pre-cut polycarbonate covered with polyester fiber mats. The fiber mats compress when inserted and "self-hold" to separate settling sand on either side.
- Pipette filler bulb attached to a steel pipe to smooth layered surfaces with gentle water puffs.
- A large diameter stiff tubing with a funnel on top for accurate direction of the sand.

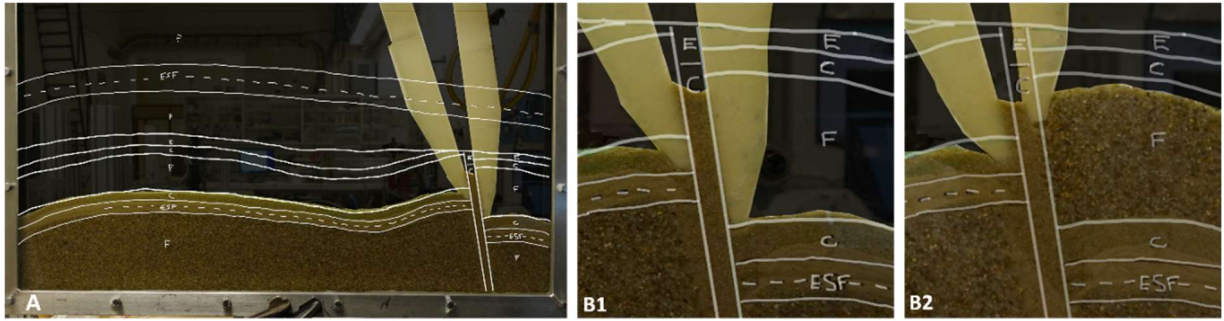

*Fig. 5: Example of how to build features in the Medium FluidFlower using unconsolidated sand. A: The geological geometry is drawn on the transparent front panel. In this example the sand adjacent to the fault is different from within the fault, and two "angle-tools" (white) separate the sands. B: Details from fault construction. The fault and adjacent facies are deposited in a layer-by-layer approach for each sand type to match planned geometry. B1 show C-sand added to the left of the fault zone before F-sand on the right (B2). The angle tool represents the boundary of the fault, and the sand to the right of the fault replaces the angle tool as it is lifted to maintain the sand within the fault.*

Key learnings: Difference between grain sizes in adjacent layers should not be too large to avoid mixing as fines fall into the coarser sand. Horizontal layers may be adjusted using a ladle constant-velocity horizontal movements along the whole unit length (especially important for fine-grain sand with longer settling time). Faults, depending on the fault design (with or without fault zone), require a minimum of one "angle tool" to separate the sand in adjacent layers.

### 2.4 Fluid injection protocols and initial conditions

The protocol to prepare for $CO_2$ injection experiments used the following steps:

1. Inject the preferred pH-indicator mix (**Table 4**) according to the experimental protocol, using the three injection ports (F1 – F3) along the bottom of the rig (**Fig. 2**).
2. Initiate inflow with open overflow port (**Fig. 2**) to ensure constant hydrostatic pressure during the experiment.
3. Bleed tubing/valves for $CO_2$ injection with fluid from the rig (and collect samples for pH measurements). Stabilize overflow before continuing.
4. Run a warm-up MFC sequence (100 % open valve for 30 min) to reduce rate fluctuations.
5. At a low $CO_2$ injection rate, connect tubing to valve and let it pressurize according to protocol before the valve is opened.
    a. Some backflow of pH-indicator is expected; record time when the injected $CO_2$ displace the backflow fluid into the valve.
    b. Continue $CO_2$ injection as described in the protocol (**Table 5**).
    c. Log inflow and outflow rate (mass per timestep) to determine displaced volume.
6. Inject degassed lye solution to reset the fluids in the porous media

Key learnings: $CO_2$ injection rate should follow a scripted MFC protocol, with sufficiently high injection rate during ramp up to maintain $CO_2$ as a gas phase in the initial stages of injection.

All experiments utilize two ports for $CO_2$ injection (one port for AC06 and AC07) at room temperature (approximately 23 °C) and atmospheric pressure at the free water table at the top of the rig (cf. **Table 6**). Temperature fluctuations were minimized, but not eliminated, in the laboratory space, and add some uncertainty.

*Table 5: Condensed $CO_2$ injection protocol used in geometry A and geometry B. Comprehensive protocol found in **SI.table 1**. I1 and I2 refer to injection in flow port 1 and port 2 respectively (**Fig. 4**).*

|  | Albus geometry | | Bilbo geometry | |
|---|---|---|---|---|
| **Objective** | Rate [ml$_s$/min][a] | Duration [hh:mm:ss] | Rate [ml$_s$/min][a] | Duration [hh:mm:ss] |
| I1 rate ramp-up | 0.1 - 1.5 | 00:05:00 | 1.0 - 1.5 | 00:02:00 |
| **I1 injection** | **2.0** | **00:45:00** | **2.0** | **01:03:13** |
| I1 rate ramp-down | 1.5 - 0.1 | 00:05:00 | 1.5 - 0.5 | 00:03:00 |
| I2 rate ramp-up | 0.1 - 1.5 | 00:05:00 | 1.0 - 1.5 | 00:02:00 |
| **I2 injection** | **2.0** | **01:15:00** | **2.0** | **01:12:52** |
| I2 rate ramp-down | 1.5 - 0.1 | 00:04:42 | 1.5 - 0.5 | 00:03:00 |

[a] $CO_2$ injection rate in ml$_s$/min @ standard conditions: 20 deg °C and 1013 millibar

*Table 6: Initial conditions for CO₂ injection experiments in the Albus (AC) and Bilbo (BC) geometries*

| Experiment | pH-indicator mix[a] | $CO_2$ inj. 1st/2nd port [g] | P [mbar] [b] 48h avg. ± σ | $CO_2$ inj. started | State of formation water degassing |
|---|---|---|---|---|---|
| AC02 | 1 | 0.179/0.283 | 999 ± 5 | Oct 13th, 2021 | Insufficient |
| AC03 | 1 | 0.176/0.283 | 994 ± 12 | Oct 18th, 2021 | Insufficient |
| AC04 | 1 | 0.176/0.283 | 996 ± 11 | Oct 21st, 2021 | Insufficient |
| AC05 | 1 | 0.176/0.283 | 995 ± 2 | Oct 26th, 2021 | Insufficient |
| AC06 | 1 | 0.459/0 | 989 ± 3 | Oct 29th, 2021 | Insufficient |
| AC07 | 1 | 0.981/0 | 993 ± 2 | Dec 7th, 2021 | Insufficient |
| AC08 | 1 | 0.176/0.284 | 1015 ± 2 | Apr 21st, 2022 | Insufficient |
| AC09 | 2 | 0.176/0.284 | 1010 ± 2 | May 3rd, 2022 | Insufficient |
| AC10 | 3 | 0.176/0.284 | 1018 ± 6 | May 6th, 2022 | Sufficient |
| AC14 | 1 | 0.176/0.284 | 1003 ± 1 | Jun 10th, 2022 | Sufficient |
| AC19 | 3 | 0.176/ 0.284 | 1020 ± 3 | Oct 18th, 2022 | Sufficient |
| AC22 | 3 | 0.176/ 0.284 | 1000 ± 1 | Nov 15th, 2022 | Sufficient |
| BC01 | 1 | 0.240/0.275 | 1020 ± 4 | May 7th, 2022 | Sufficient |
| BC02 | 1 | 0.240/0.275 | 992 ± 3 | May 23rd, 2022 | Sufficient |
| BC03 | 2 | 0.240/0.275 | 1006 ± 2 | Aug 17th, 2022 | Sufficient |
| BC04 | 3 | 0.240/0.275 | 1019 ± 3 | Aug 31st, 2022 | Sufficient |

[a] Ref Table 4
[b] Atmospheric pressure from metrological data in Bergen (Norway) during each cycle **(SI.fig. 4)** (Geophysical Institute, 2022).

## 2.5    Image acquisition and analysis

*Camera settings*

Because the two rigs were operated in parallel, two different cameras were used. For the Albus geometry the camera (Sony ZV-1) used the following settings: exposure time 1/4 sec, F number 2.5, ISO 200, and manual focus. The camera was positioned in front of the rig. Images were captured at 5472x3080 pixels every 20 or 30 sec during $CO_2$ injection, and every 300 sec afterwards until 48 hours after $CO_2$ injection was initiated. While for the Bilbo geometry, the camera (Sony A7III, lens SAMYANG AF 45 mm F1.8) used the following settings: exposure time 1/6 sec, F number 2.2, ISO 100, and manual focus. Images were captured at 7952x4472 pixels every 30 sec during $CO_2$ injection and every 300 sec afterwards until the end of the experiment at 48 hours.

Because of the image interval used to capture dynamics during each experiment, the full dataset consists of ~4800 images for each experiment. A subset that captures most of the flow dynamics was generated and used for the analysis presented here. These subsets consist of 214 images with the following intervals: 10 images prior to $CO_2$ injection start, images every 5 min during the first 360 min (6 hours), images every 10 min from 360 min until 1440 min (24 hours), and images every 60 min from 1440 min until end of experiment at 2880 min (48 hours).

*Phase segmentation through image analysis*

The high-resolution images, acquired as described above, allow for a detailed visual examination of the fluid displacement along the front panel; fluid flow inside the reservoir remains unnoticed, yet the low depth of the geometry suggests relatively small deviations averaged over time. Micro scale features such as gas bubbles or sand grain motion as well as macro scale features as the phase segmentation of the $CO_2$-water mixture into the phases of water, $CO_2$ saturated water, and $CO_2$ gas can be identified from the photographs. The latter is possible due to the use of pH sensitive dye, cf. **Fig. 3**. To analyze the large and varying dataset for all experiments, the image analysis toolbox DarSIA (Nordbotten et al., this issue) is used. Altogether, the combination of photographs and image analysis software constitutes one of the main measurement instruments used in this work.

Prior to any analysis, DarSIA is used to unify the environment including the aligning of images and restricting to a fixed region of interest, normalizing temporal color and illumination fluctuations, as well as monitoring and correcting for small sand settling events. Having a cleaned representation of each image allows for comparing it to a fixed baseline image and therefore tracking advancing fluids as differences to the baseline.

To quantitatively describe the multiphase flow in the Medium FluidFlower rig, a range of assumptions must be made. These are the same assumptions as used by Fernø et. al (this issue) to analyze multiphase flow images from the large FluidFlower rig, and a summary is included below:

I. we assume that gas-filled regions are 100% saturated with the gas ($CO_2$)
II. we assume a constant $CO_2$ concentration in the $CO_2$ saturated water
III. we do not account for the dynamics of the gas partitioning in the gas accumulation
IV. we can accurately calculate the volume of $CO_2$ injected during each run
V. we have accurate information about porosity and depth

Based on these assumptions, a tertiary phase segmentation of the images, identifying the formation water, dissolved $CO_2$ and mobile $CO_2$ phases, also locates the presence of all $CO_2$ within the rig. The algorithmic phase segmentation boils down to thresholding both in terms of color and signal intensity. Depending on the pH-indicator used, two suitable monochromatic color channels are picked aiming at identifying first all $CO_2$ within the water, and then the gaseous $CO_2$ within all $CO_2$. Each sand type, saturated or unsaturated, reflects light differently. Thus, the thresholding algorithm is designed to take into account the heterogeneous nature of the geometry. It automatically dissects the analysis into sub-analyses of the different sand layers including choosing dynamic thresholding parameters for the different phases and layers. With this, an accurate segmentation of $CO_2$ and water is possible. The detection of gaseous $CO_2$ occurs on the Darcy scale. It should be emphasized that this possibly leads to ignoring single gas bubbles while at other times enlarging them due to the averaging procedure and specific choice of thresholding parameters, used for converting the fine scale images to coarse scale data. On the larger scale, this effect is noticeable but plays only a minor role. The segmentation and its accuracy are illustrated in **Fig. 6**.

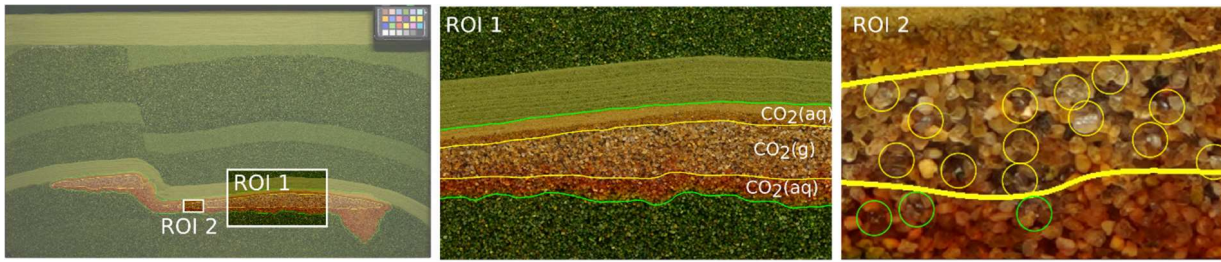

*Fig. 6: Left: Illustration of the phase segmentation for BC01 after one hour of $CO_2$ injection. Center (ROI1): Segmentation with the transition zones of mobile to dissolved $CO_2$ as well as from dissolved $CO_2$ to water being entirely associated to the mobile and dissolved $CO_2$ phases, respectively. Right (ROI2): Highlighted, manually detected gas bubbles associated to dissolved (green circles) and mobile (yellow circles) $CO_2$ phases, illustrating the accuracy of the segmentation algorithm.*

*Sensitivity of the image analysis with respect to the pH-indicator mixes*
The different pH-indicator mixes constitute the visual markers for the image analysis, and threshold parameters have to be chosen to properly identify the different phases. As seen from Fig. 3, the responses of the different pH-indicators to the presence of $CO_2$ span different ranges of color variations. Comparing the chemically similar pH-indicator mixes 1 and 2, the first shows a larger span of colors than the latter. As a consequence of the latter, the thresholding algorithm is more sensitive with respect to the threshold parameters, resulting in systematic over-detection of the gaseous phase (**Fig. 7**). A similar conclusion can be drawn for the comparison of the chemically different pH-indicator mixes 1 and 3. Across the different experiments using the same pH-indicator mix, the same thresholding algorithms and parameters have been used. The calibration of the three sets of parameters has been however performed purely based on visual examination and without comparison across the mixes.

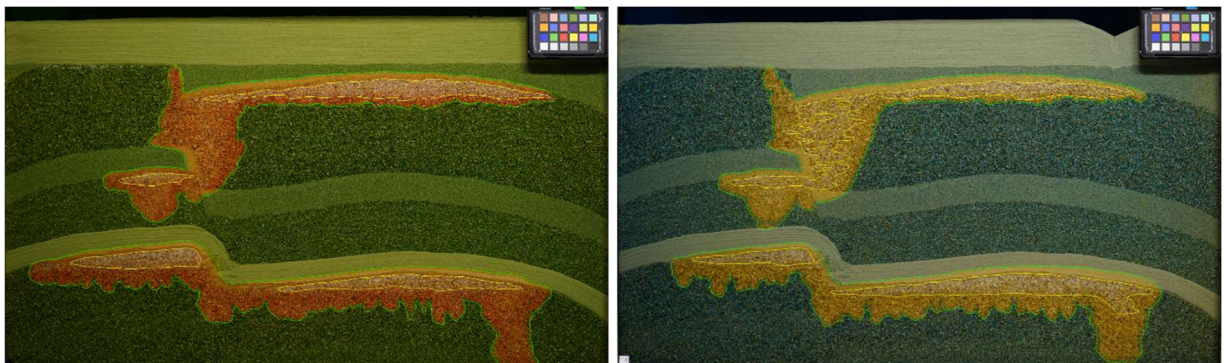

*Fig. 7: Phase segmentation for the same experimental setup under the same conditions, only differing the use of the pH indicator mix. Left: PH-indicator mix 1 is used, and the segmentation detects gaseous regions (yellow contours) based on variations in the red color and tuned to detect dry sand. Right: PH-indicator mix 2 is used, and the segmentation detects*

gaseous regions (yellow contours) based on the variations in the yellow color, which are less pronounced compared to the left image. As a result, residual trapped and rising mobile $CO_2$ is detected in the upper zone, on the right.

*Quantitative analysis of physical variability*
To compare different experiments, the phase segmentations are utilized by DarSIA to provide visualizations as in **Fig. 12.** There, different colors are assigned to appearances of different segmentations and their overlaps. Moreover, the fractions that each color represents in reference to the total covered area are calculated (and printed in the legends). In **Fig. 12**, the depth map of the rig (**SI.fig. 5**) is also considered when computing the fractions such that it really becomes volume fractions and not area fractions.

*Mass calculations*
After an image is segmented into formation water, $CO_2$-saturated water and mobile $CO_2$, the total mass of each $CO_2$ phase can be determined. The total $CO_2$ mass $m_{CO_2}^{total}$ is known at any time, cf. injection profile in **Table 5**. Thus, assuming that all $CO_2$ in the rig either is mobile or dissolved, cf. assumption III, it is sufficient to determine the mass of mobile $CO_2$, $m_{CO_2}^{mobile}$, while the mass of dissolved $CO_2$ is given by $m_{CO_2}^{dissolved} = m_{CO_2}^{total} - m_{CO_2}^{mobile}$.

The mass of mobile $CO_2$, $m_{CO_2}^{mobile}$, is determined as pixel-wise sum of the pixel-wise defined mass density of mobile $CO_2$, $\rho_{CO_2}^{mobile} = \phi \cdot d \cdot A \cdot s_g \cdot \chi_{CO_2}^{g}$. Here, $\phi, d, A$ denote the local porosity and depth as well as the pixel area, respectively, constituting together the local pore volume, which according to assumption V can be determined accurately, cf. also **SI.fig. 4**. Based on assumption I, the saturation $s_g$, takes the value 1 in the region of detected mobile $CO_2$ and 0 otherwise and is thereby fully prescribed by the phase segmentation. It remains to identify the mass concentration of $CO_2$ in the gaseous phase $\chi_{CO_2}^{g}$ which is given by the density of gaseous $CO_2$ under operational conditions, obtained from the NIST database (Lemmon et al.2022) Here, a uniform temperature distribution of 23 deg Celsius is assumed in the rigs; in addition, the fluid pressure is determined from local meteorological weather measurements, cf. **SI.fig. 2**, taking into account the local height difference between the FluidFlower rigs (~29 m corresponding to increase of 3.625 mbar) and additional hydrostatic pressure of 1013.25 mbar/m from the free water level. This finally determines $m_{CO_2}^{mobile}$.

## 2.6   Experimental results and variations in $CO_2$ migration patterns

Here we compare and discuss the twelve $CO_2$ injection experiments in the Albus geometry, and the four $CO_2$ injection experiments in the Bilbo geometry. Visual observations from $CO_2$ injections in each geometry are described, and experiments are discussed with regards to utilizing the Medium FluidFlower rigs for rapid evaluation of key operational conditions, including the effect of degassed aqueous phase and different pH-indicator mixes.

*$CO_2$ migration patterns in the Albus and Bilbo geometries*
The $CO_2$ migration for 10 operationally comparable $CO_2$ injection experiments in the Albus geometry follows a similar pattern, which is described next with reference to **Fig. 8**: The gaseous $CO_2$ injected in the *Lower F* layer (I1) quickly dissolves in the formation fluid and changes the color of the pH-indicator. With continued $CO_2$ injection, the $CO_2$ saturated water spreads out in a U-shape from the injection port, upwards until it reaches the *Lower seal*. At this stage both gaseous $CO_2$ and $CO_2$ saturated water are visible. $CO_2$ migrates to the left below the *Lower seal* and accumulates in the anticlinal fold trap (contour 1 – light blue). As the gas accumulation increases in the trap, the U-shape above the injection port expand, and gravitational fingers develop under the gas accumulation when $CO_2$ injection in port 1 stops (contour 2 – dark blue). The migration from the second $CO_2$ injection port (I2) is initially characterized with a small gas accumulation in the small anticlinal trap below the *Upper C* layer on the left side of the fault (contour 3-light green). When the gas reaches the anticlinal trap spill point (**Fig. 4**, top), buoyancy forces cause the gas to continue through the *Upper E* layer and upwards until it reaches the *Top Regional seal* (contour 4-dark green). The gas migrates stepwise upwards under the sealing unit and sequentially fill the anticlinal trap with gas. Meanwhile, the gravitational fingers below the lower gas accumulation grow (contour 2-5); after the second injection ceased (contour 5-pink) fingers develop below the *Top Regional seal* (contour 6-red). The fingers moved laterally when reaching the *Upper C* layer, before some fingers eventually continue downwards in the *Middle F* layer below.

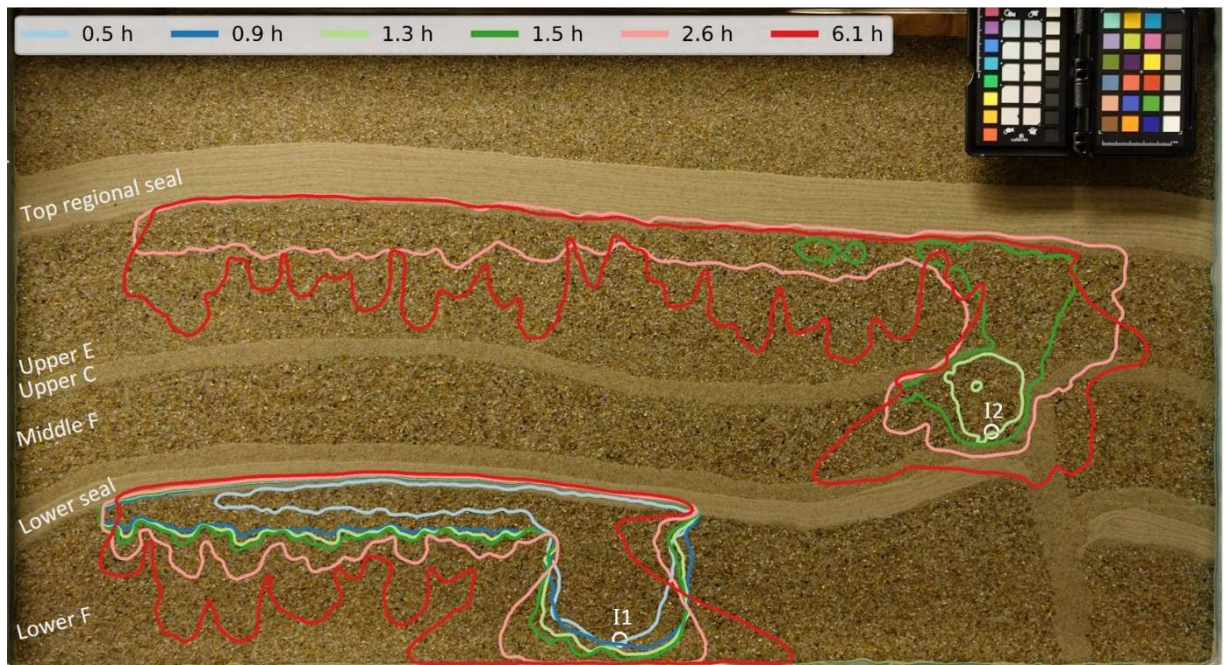

*Fig. 8: The migration pattern of $CO_2$ during experiment AC10, representing the general pattern observed for all 10 experiments (AC01 – AC10) in the Albus geometry. The contours represent the distribution of gaseous and aqueous $CO_2$ at different times: 0.9 hours represent end of first injection (I1) and 2.6 hours represent end of second injection (I2).*

The $CO_2$ migration for four $CO_2$ injection experiments in the Bilbo geometry follows a similar pattern (differences will be discussed later in this section), see **Fig. 9**: The gaseous $CO_2$ injected in the *Middle F* layer (I1), and the U-shape described for the Albus geometry emerges also here. Both residually trapped $CO_2$ bubbles and $CO_2$ saturated water occur above the injection point, with gas migrating to the left when reaching the *Lower seal* unit (contour 1 - light blue). Residually trapped gas bubbles are observed from the injector to the front of the advancing gas as it fills the anticline trap. The gas accumulation slightly exceeds the spill-point (contour 2 – dark blue) before it "burst" leftwards into the smeared fault trap, a process that is repeated multiple times (see discussion in Fernø et al., this issue). Some residually trapped gas bubbles are observed in the vicinity of the "smeared" *Lower seal* area, and a new gas accumulation develops in the smeared fault trap when port 1 $CO_2$ injection stopped (contour 3- light green). The $CO_2$ injection continues in the second port (I2), located in the hanging wall of the *Upper E* layer. A small gas accumulation is observed before the gas exceeds the spill-point below the hanging wall of the faulted *Middle C* layer above (contour 4 – dark green). Some gas flow left of the fault plane and into the hanging wall of the *Upper F layer,* but most of the $CO_2$ migrates right and accumulates under the anticline trap in the footwall of the *Upper F* layer. As seen in the Albus geometry, after injection has ceased in the Bilbo geometry (contour 5 - pink), gravitational fingers of $CO_2$-saturated water develop and sink downwards (contour 6 - red).

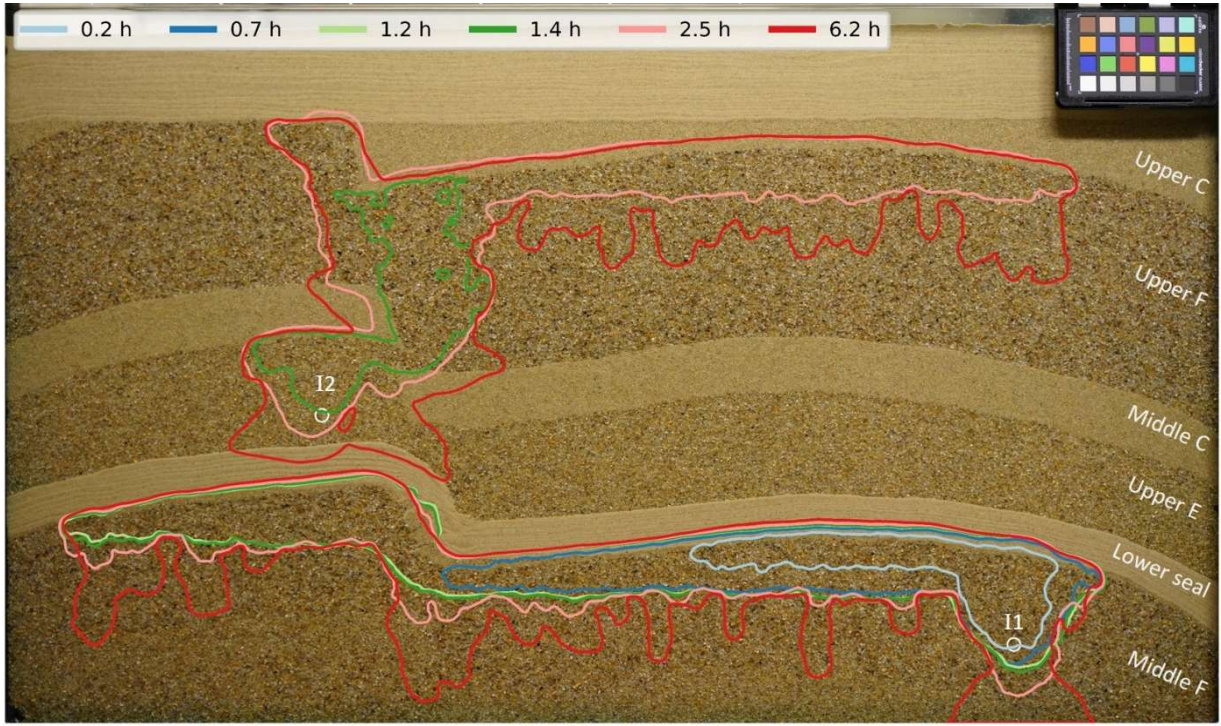

*Fig. 9: The migration pattern of $CO_2$ during experiment BC02, representing the general pattern observed for all four (BC01-04) experiments in the Bilbo geometry. The contours represent the distribution of gaseous and aqueous $CO_2$ at different times: 1.2 hours represent end of first injection (I1), and 2.5 hours represent end of second injection (I2).*

*Impact from degree of degassed aqueous phase*

$CO_2$ injections in the Albus geometry illustrate the impact from variable degassing of the formation water. Of the 10 experiments with the same injection protocol, six experiments have pH-indicator mix 1 (**Table 6**). Development of calculated mass of mobile $CO_2$ and dissolved $CO_2$ over time show a wide spread of the results (**Fig. 10**). The effect of insufficient degassing is evident for the mobile gas at later times: with sufficient degassing, the mass of mobile CO2 is zero for later times (AC14) because all of the injected $CO_2$ is dissolved in the formation water. In contrast, with insufficient degassing (atmospheric gases present in the aqueous phase) a gaseous phase remains in the geometry and is included in the mass calculations of mobile $CO_2$. The remaining gas at late times (observed in AC02-05 and AC08) is decreasing amounts of air, and not $CO_2$.

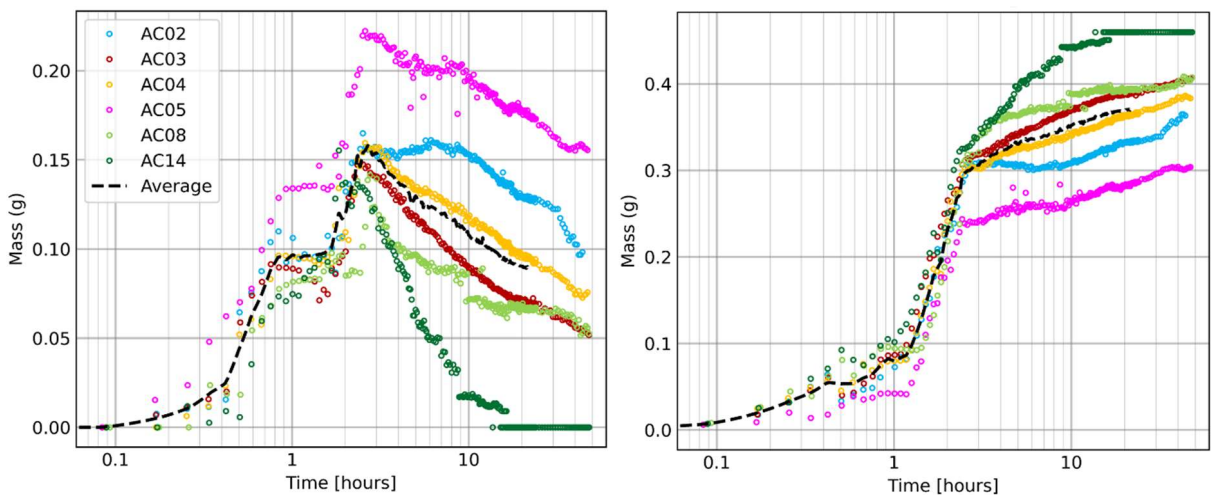

*Fig. 10: Comparison of $CO_2$ injection experiments in the Albus geometry with the same injection protocol and pH-indicator mix. Left plot: development of calculated mass of mobile $CO_2$ over time, and right plot: development of calculated mass of dissolved $CO_2$.*

The degree degassing influences the distribution of mobile and dissolved $CO_2$ in Albus geometry. Nevertheless, the overlap in spatial distribution of mobile and dissolved $CO_2$ for six experiments in the Albus geometry has an overall average of 65% overlapping (**Fig. 11**). Operational inconsistencies that reduce the overlap include 1) some $CO_2$ was injected in AC02 prior to starting the experiment (ramp-up), and 2) $CO_2$ injection was not scripted in the first experiments, causing deviation in time between $CO_2$ injection in port 1 and port 2 **(SI.table 1)**. The spatial distribution are included in supplementary information (**SI.fig. 6**).

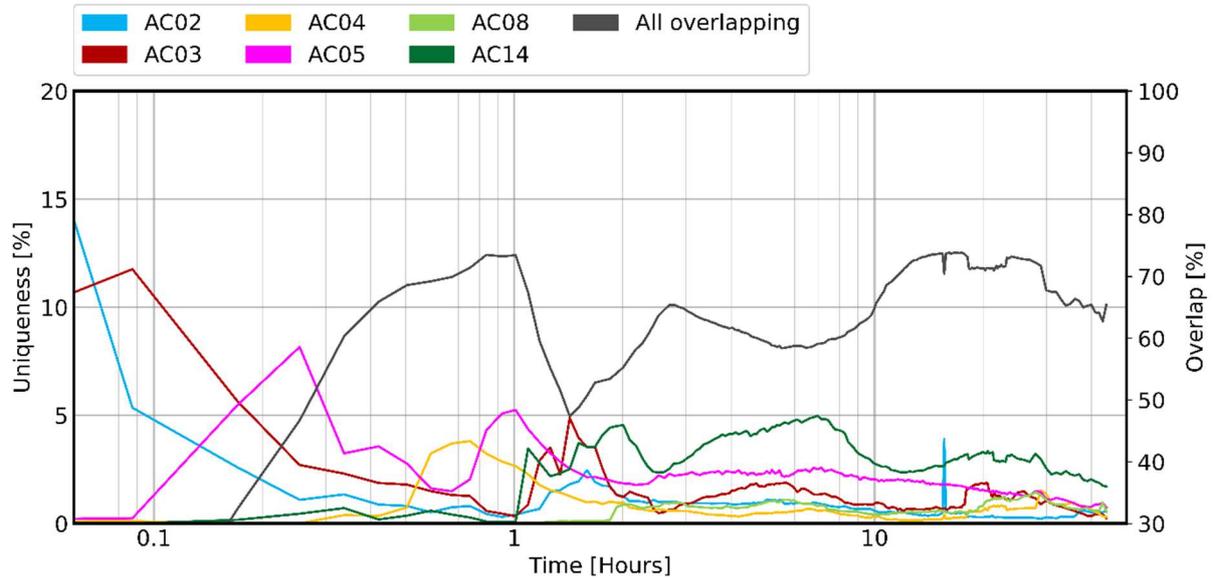

*Fig. 11: Development of uniqueness of phase segmentations of dissolved $CO_2$ for six $CO_2$ injection experiments in the Albus geometry. Most of the injections have less than 5 % uniqueness. The degree of overlap between all six experiments (gray, secondary y-axis) reaches a maximum (approximately 75 %) after one hour, with an average of 65 % for the duration of the experiments. Note that the x-axis is logarithmic.*

*Impact from different pH-indicator mixes*

Three different pH-indicator mixes are used in the Bilbo geometry, varying in parts in their chemical properties (section 2.1, fluids), as well as their interaction with the image analysis (detailed in section 2.5). The development in uniqueness and overlap (**Fig. 12**, top) demonstrate that the experiments are comparable with less than 5 % uniqueness for most of the time, where changes originate from variations in atmospheric pressure (**SI.fig. 4**) or methylene red precipitation (**SI.fig. 7**). As expected, the gas dissolves faster with less spreading of dissolved $CO_2$ in the BC04 experiment (pH ~10.4) compared to the experiments BC01– 03 (pH ~8.3), which is reflected in the low overlap of "BC01+BC04" compared to "BC01+BC03" (**Fig. 12**, top). Furthermore, the observation is supported by the evolution of the masses of the different $CO_2$ phases (**Fig. 13**). In the same figure, a clear difference between BC01 and BC03 can be observed which is directly connected to the sensitivity of the image analysis and the detection of mobile and residual trapped gaseous phase, as discussed in section 2.5.

Similar material for $CO_2$ injection experiments in the Albus geometry, comparing results from pH-indicator mix 1-3 are included in supplementary information (**SI.fig. 8** and **SI.fig. 9**). To summarize, the results follow the same trends as for the $CO_2$ injection experiments in the Bilbo geometry, as presented above. We highlight the comparison of pH-indicator mix 1 and 2 (AC14 and AC09), showing an average overlap of only 81% (**SI.fig. 9**), which again is attributed to the interplay of the mixes as visual markers and the image analysis. Furthermore, for the two experiments with higher degree of vacuuming of initial fluids (AC19 and AC22), an increase of the dissolution rate results in earlier attaining zero mobile/residual gas compared to the other experimental runs (**SI.fig. 8**).

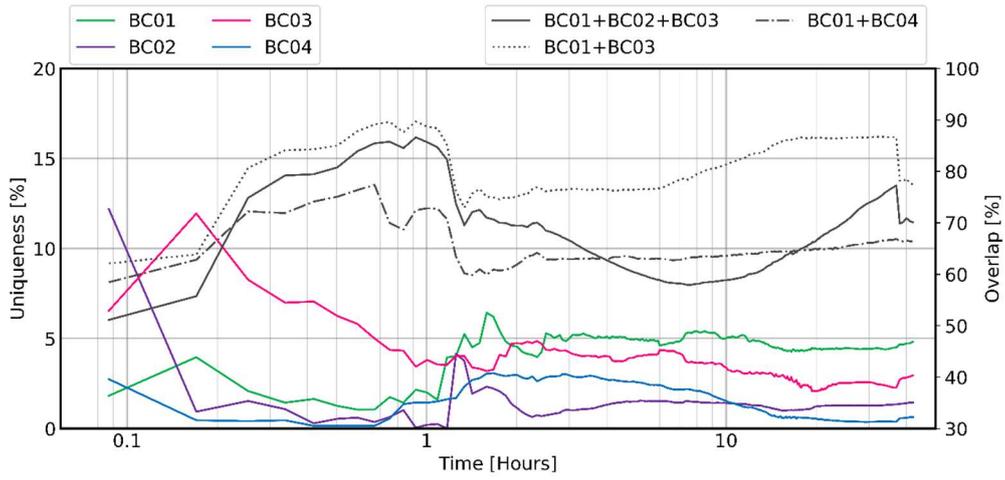

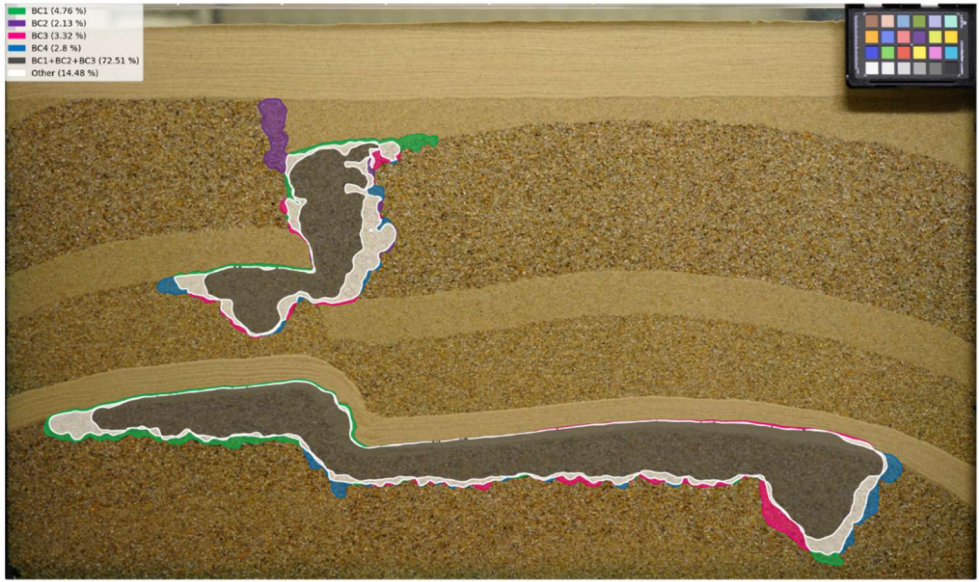

*Fig. 12: The development in uniqueness and overlap for experiments BC01-04. Top: quantitative uniqueness for BC01-04 and overlap for combinations of experiments. Bottom: Spatial uniqueness after 1.5 hours in BC01 (green), BC02 (purple), BC03 (pink) and BC04 (blue); overlap for BC01-03. Other overlap combinations are lumped together (white). Each color represents the spatial distribution of mobile and dissolved $CO_2$. Additional timesteps are shown in **SI.fig. 7**.*

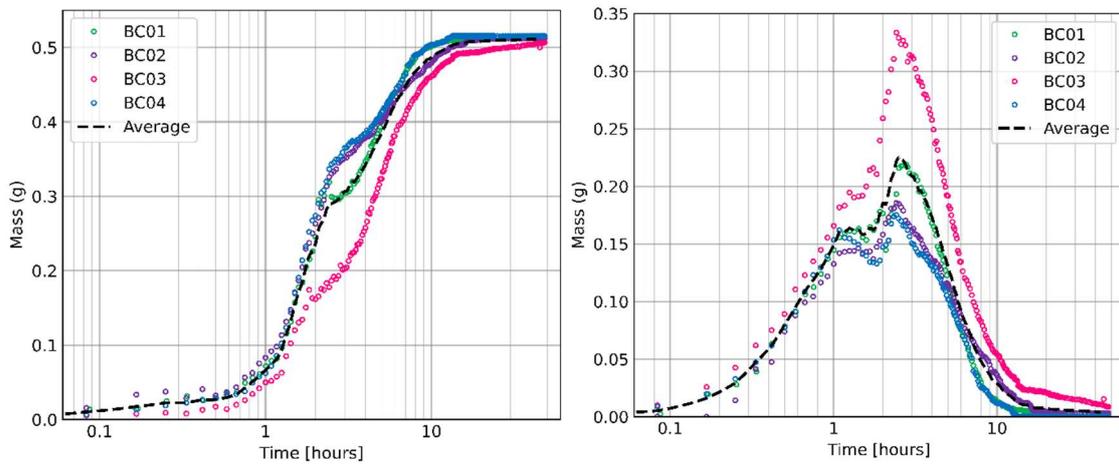

*Fig. 13: Development in mass [g] of dissolved $CO_2$ (left) and mobile $CO_2$ (right) for four $CO_2$ injection experiments (BC01-04) in the Bilbo geometry. The overall behavior of dissolved $CO_2$ is comparable for all experiments, the deviation of BC03 relates to sensitivity of the image analysis with respect to threshold parameters (discussed in Section 2.5). Experiment with pH-indicator mix number 3 (BC04) has faster $CO_2$ dissolution compared to pH-indicator mix number 1 and 2.*

## 3 Numerical modeling of experimental $CO_2$ injection

Safe geologic storage of $CO_2$ requires numerical modeling for forecasting $CO_2$ migration in complex geological structures. The computational models used are typically physics-based, and are expected to include all dominant processes. Nevertheless, the accuracy of these models is hard to quantify due to lack of direct observations in field conditions. The FluidFlower benchmark study (Flemisch et al, this issue), provides an insight into the accuracy of numerical models for $CO_2$ storage. Furthermore, Saló-Salgado et al. (this issue) aims to refine our understanding of the accuracy of numerical models by systematically evaluating the value of increasing amount of local data in predicting the $CO_2$ migration in FluidFlower experiments. This provides a unique opportunity to evaluate both the measurements of the petrophysical properties of the unconsolidated sand (section 1.4) and the development during the $CO_2$ injection experiments in the two geometries built in the Medium FluidFlower rigs (section 2).

Saló-Salgado et al. (this issue) presents three different models (denoted I through III), with respectively increasing amount of experimentally measured petrophysical data, while keeping all other model characteristics constant:
- Model I considers only grain size width as provided by the experimental measurements (**Table 2**) to estimate petrophysical data from published data on similar silica sand. History matching this model to the AC02 experiment required 10 iterations, where parameters were successively updated manually.
- Model II furthermore considers single-phase data from the experimental measurements, while the multiphase data remains based on published data. History matching this model to the AC02 experiment required 8 runs.
- Model III uses all the data provided from the experimental measurements (**Table 3**), and required 7 iterations to match the results in AC02.

Evaluation and comparison of the modeled $CO_2$ migration and the direct observations from the AC02 experiment in the Albus geometry is presented in Saló-Salgado et al. (this issue). To test the robustness of the simulation model, the same petrophysical properties as obtained from history match in the respective models (I - III) are used to predict migration development during $CO_2$ injection in Bilbo geometry (**Fig. 4**), and this is detailed below.

### 3.1 Model properties used for the geometry in the Bilbo rig

*Numerical model setup*

Simulation results presented here are performed with the black oil module in the Matlab Reservoir Simulation Toolbox (MRST) (Lie, 2019), where properties of the water are assigned to the oleic phase. Structural trapping, dissolution trapping and residual trapping (Juanes et al. 2006) is included, and details about implementation of these can be found elsewhere (Saló-Salgado et al., this issue). Due to the buoyancy of $CO_2$ at atmospheric conditions and high permeability of the sand, very small timesteps are required in general, however the main numerical difficulty is nevertheless related to convergence of the non-linear solver. PVT properties used in the simulations are according to atmospheric conditions (25 deg C), with $CO_2$ in gaseous phase. The thermodynamic model is the same as applied in Saló-Salgado et al. (this issue).

Dimensions of computational grid used to model the porous medium in the Bilbo rig are: 93.4 x 53.3 x 1.05 cm. Layer contact coordinates are extracted from a 2D image, and the grid has a cell size of 5mm and a total of 20,470 cells, with single cell layer to account for thickness and volume and obtain a 3D grid (**SI.fig. 10**). There are no-flow boundary conditions everywhere except at the top surface, where there is constant pressure consistent with atmospheric pressure and a fixed height water table. Initial conditions are steady state, consistent with the boundary conditions. Injection is conducted through ports (1.8 mm diameter) completed in a single cell at the respective coordinates (**Fig. 4**). In a water filled porous medium, $CO_2$ injection schedule in the simulations is the same as in the Bilbo experiments (**Table 5**).

*Petrophysical input parameters*

The input values for porosity, absolute permeability and relative permeability used in the Bilbo model are identical to the ones used to match the AC02 experiment for model I to III (**SI.table 2**), detailed in Saló-Salgado et al. (this issue). During $CO_2$ injection in AC02, the sealing capacity of the C-sand was not critical due to the location of the injection ports and the injection protocol used. Hence, the calibration based on history matching this injection retain a high degree of uncertainty. This becomes prominent when simulating the Bilbo-geometry, where the second injection port is in the *Upper E* layer, overlayed by the faulted *Middle C* layer (**Fig. 4**). When the Albus AC02 HM values are used for gas saturation and capillary entry pressure for sand C in the Bilbo geometry, free-phase $CO_2$ and $CO_2$ saturated water migrate through the hanging wall of the *Middle C* layer instead of filling the fault trap to the spill-point. To mitigate this effect, we choose to decrease the endpoint gas saturation (Sg) and increase the capillary entry pressure (Pc) in the C-sand (**Table 7**). The increased capillary entry pressure for the vertical fault zone was also observed experimentally (see **SI.fig. 3**).

Table 7: C-sand values for endpoint gas saturation (Sg) and capillary entry pressure (Pc) obtained from history match of $CO_2$ injection experiment in the Albus geometry, compared to values needed to match $CO_2$ injection experiment in the Bilbo geometry. All other parameters kept the same as obtained from HM of the Albus experiment for the three models.

| Model | Albus HM: C-sand | | Bilbo HM: C-sand | |
|---|---|---|---|---|
| | Sg | Pc [mbar] | Sg | Pc [mbar] |
| I | 1e-3 | 4.6 | 1e-4 | 9.3 |
| II | 1e-3 | 2.6 | 1e-4 | 10.0 |
| III | 1e-3 | 4.5 | 1e-4 | 9.6 |

## 3.2 Simulation results

When evaluating the simulation results there are known deviations between the setup of the numerical simulations and the physical geometry that should be kept in mind: 1) discrepancy in temperature value used in mass calculation from experimental results and model input cause difference in total injected mas (**SI.fig. 11**), and 2) variations in flow cell depth, where calculations from experimental results include the depth map (**SI.fig.5**) with expansion of up to ~40%, while in the model a constant expansion of 5% is used.

A qualitative comparison of gas saturation and concentration of dissolved $CO_2$ shows that all three models provide similar, and fairly accurate, results for the $CO_2$ migration during $CO_2$ injection experiment in the Bilbo geometry (**Fig. 14**). The relative differences between the models appear to be comparable to the difference between the model and the experiment. However, there are subtle differences:

- After 3 hours, model I shows $CO_2$ in the hanging wall of the *Upper F* layer (top left corner in **Fig. 14**) equal to the experiment. However, this is due to $CO_2$ spilling out of the anticline from the right, rather than pore variability and diverging path possibilities observed experimentally.
- The models show $CO_2$ concentration reaching higher elevation within the middle-left and top-right C sands, with respect to the experiment.

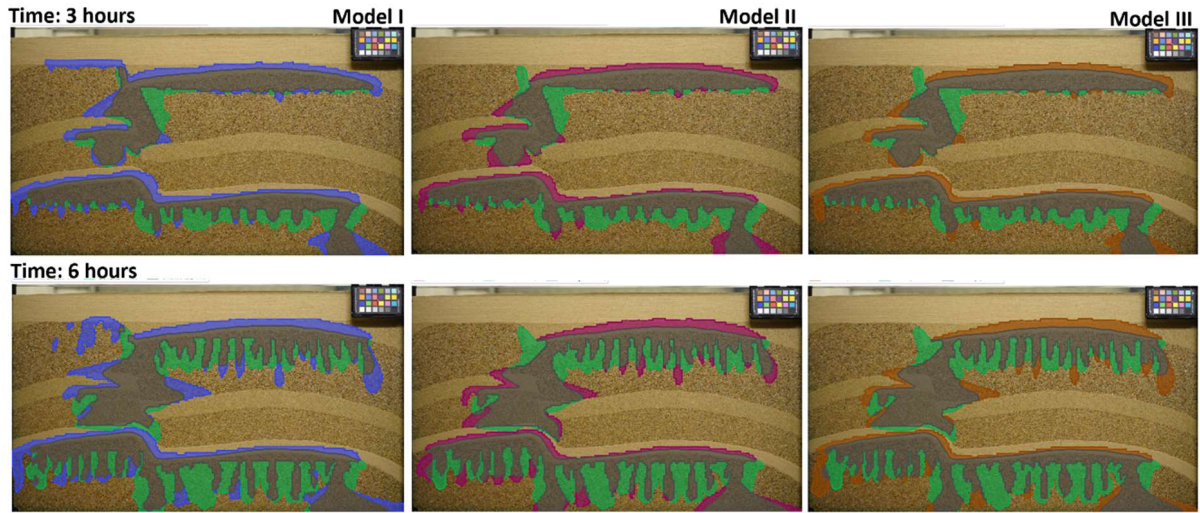

Fig. 14: Spatial uniqueness after 3 and 6 hours for Model I (blue) Model II (dark pink), Model III (brown) and experiment BC01 (green); overlap between each model and experiment in grey. Each color represents the spatial distribution of mobile and dissolved $CO_2$. Contours of the simulation results are obtained from gas concentration maps, and a threshold value of 0.21 kg/m3 (15% of the maximum value around 1.4 kg/m3) is used.

The spatial distribution of mobile $CO_2$ (gas phase) and dissolved $CO_2$ over time (**SI.fig.12**) is quantified in **Table 8,** where the unique (for each case) and overlapping phase segmentations for the three models (I, II, III) and experiment BC01 is compared. Note that lower uniqueness in **Table 8** indicates better match with experiment. The distribution of mobile $CO_2$ for Model II and III overlaps with BC01, demonstrated as zero uniqueness for time steps in **Table 8**.

*Table 8: Model-experiment comparison of the unique and overlapping spatial distribution of mobile and dissolved $CO_2$. Numbers are % of all regions with $CO_2$.*

| Time [hours] | Mobile $CO_2$ [a] | | | | | Dissolved $CO_2$ [b] | | | | |
|---|---|---|---|---|---|---|---|---|---|---|
| | Uniqueness | | | | Two or more overlap | Uniqueness | | | | Two or more overlap |
| | Model I | Model II | Model III | BC01 | | Model I | Model II | Model III | BC01 | |
| 3 | 11.4 | 0.0 | 0.0 | 4.3 | 84.2 | 7.4 | 0.9 | 0.8 | 13.9 | 77.0 |
| 6 | 28.1 | 0.0 | 0.0 | 0.5 | 71.4 | 6.6 | 1.1 | 1.3 | 19.2 | 71.9 |
| 9 | 31.2 | 0.0 | 0.0 | 0.2 | 68.6 | 4.1 | 0.9 | 0.7 | 18.4 | 75.9 |
| 12 | 38.9 | 0.0 | 0.0 | 0.4 | 60.7 | 3.4 | 0.8 | 0.3 | 17.1 | 78.4 |
| 18 | 66.8 | 0.0 | 0.0 | 0.0 | 33.2 | 2.8 | 0.4 | 0.2 | 13.4 | 83.2 |
| 24 | 71.1 | 0.0 | 0.0 | 0.0 | 28.9 | 2.3 | 0.4 | 0.3 | 11.1 | 85.9 |

[a] For simulations, gas saturation of 0.001 was used as threshold during segmentation with DarSIA
[b] For simulations, concentration of 0.21 kg/m$^3$ (15% of the maximum value around 1.4 kg/m$^3$) was used as threshold during segmentation with DarSIA (numbers from **SI.fig.12**)

Overall, models II and III perform similarly, and correlate better with the experimental results than model I. This is expected based on the model definitions, and in line with the results as obtained when matching the models to $CO_2$ injection experiment in the Albus geometry (Saló-Salgado et al., this issue). However, the numerical model is missing some physics compared to the experiment. An example of this is the compact sinking front with very thick and only moderately protruding fingers seen in the experiment, whereas the model shows thinner fingers sinking from a receding front, even when the diffusion coefficient is increased. We expect that this deviation between model and experiment can be reduced by incorporating dispersion in numerical model.

## 4 Conclusions and future outlook

The physical Medium FluidFlower rig represents a fast-prototyping tool to evaluate the parameter and operational space, and has been an essential part in development and planning of room-scale experiments and the International FluidFlower Benchmark initiative. Key learnings for constructing geological geometries (using unconsolidated sand) and fluid injection protocols are included, and was utilized during construction of the geological model in Fernø et al. (this issue). The in-house developed open-source software DarSIA analyzes high-resolution images to quantify key parameters and variability in the experimentally observed $CO_2$ migration patterns. The results show anticipated behavior of injected $CO_2$, however with physical variabilities induced by design (different formation water chemistry) and because the system is sensitive (atmospheric pressure). Numerical modeling of $CO_2$ injection experiments has predicted fairly accurate results for the $CO_2$ migration and have demonstrated the value of including measured petrophysical properties of the porous media in the simulation models. Hence, the rig represents a unique possibility to test our simulation skills because here we can compare predictions to observations.

*Future outlook.* The Medium FluidFlower with its fast-prototyping nature and the presented workflows provide an excellent opportunity to address various research and particularly modeling questions. The range of possible phase configurations, combined with the quick (and if needed recyclable) setup allows for conducting a series of varying experiments and thereby perform a comprehensive physical sensitivity study, aiming at studying isolated phenomena. The access to dense observation data and a comparison with corresponding simulation data opens up for better understanding and the possibility for improved modeling.

**Acknowledgements**

The work of JWB is funded in part by the UoB Academia-project «FracFlow» and the Wintershall DEA-funded project «PoroTwin». MH is funded from Research Council of Norway (RCN) project no. 280341. KE and MH are partly funded by Centre for Sustainable Subsurface Resources, RCN project no. 331841. BB is funded from RCN project no. 324688. LS gratefully acknowledges the support of a fellowship from "la Caixa" Foundation (ID 100010434). The fellowship code is LCF/BQ/EU21/11890139.The authors would also like to acknowledge Ida Louise Mortensen and Mali Ones for their contribution in the lab during their internship at UoB.

# References

Beard, D.C. and P.K. Weyl, *Influence of Texture on Porosity and Permeability of Unconsolidated Sand1*. AAPG Bulletin, 1973. **57**(2): p. 349-369.

Chapuis, R.P., *Predicting the saturated hydraulic conductivity of soils: a review.* Bulletin of Engineering Geology and the Environment, 2012. **71**(3): p. 401-434.

Eikehaug, K., Haugen, M., Folkvord, O., Benali, B., Bang Larsen, E., Tinkova, A., Rotevatn, A., Nordbotten, J.M., Fernø, M.A. Engineering meter-scale porous media flow experiments for quantitative studies of geological carbon sequestration, TiPM SI (2023), submitted

Fernø, M.A., Haugen, M., Eikehaug, K., Folkvord, O., Benali, B., Both, J.W, Storvik, E., Nixon, C.W., Gawthrope, R.L. and Nordbotten, J.M.: Room-scale CO2 injections in a physical reservoir model with faults, TiPM SI (2023), submitted

Flemisch B, Nordbotten JM, Fernø MA, Juanes R, Class H, Delshad M, Doster F, Ennis-King J, Franc J, Geiger S, Gläser D, Green C, Gunning J, Hajibeygi H, Jackson SJ, Jammoul M, Karra S, Li J,Matthäi SK, Miller T, Shao Q, Spurin C, Stauffer P, Tchelepi H, Tian X, Viswanathan H, Voskov D, Wang Y, Wapperom M, Wheeler MF, Wilkins A, Youssef AA, Zhang Z. *The FluidFlower International Benchmark Study: Process, Modeling Results, and Comparison to Experimental Data,* TiPM SI (2023), submitted

Folk, R. L. & Ward, W. C. (1957). Brazos River Bar: a Study in the Significance of Grain Size Parameter. Journal of Sedimentary Petrology, 27: 3-27.

Geophysical Institute, University of Bergen, https://veret.gfi.uib.no/?action=download

IEA, *Net Zero by 2050*. 2021: Paris

Juanes, R., Spiteri, E. J., Orr, F. M., and Blunt, M. J. (2006), Impact of relative permeability hysteresis on geological CO2 storage, Water Resour. Res., 42, W12418, doi:10.1029/2005WR004806.

Krumbein, W. C.: Application of Logarithmic Moments to Size Frequency Distributions of Sediments. Jnl. Sed. Petrology (1936) 6,35-47.

Krumbein, W.C. and Monk, G.D., Permeability as a Function of the Size Parameters of Unconsolidated Sand. Petroleum Technology, 1942

Lemmon, E.W., Bell, I.H., Huber, M.L., McLinden, M.O., Thermophysical Properties of Fluid Systems, NIST Chemistry WebBook, NIST Standard Reference Database Number 69, Eds. P.J. Linstrom and W.G. Mallard, National Institute of Standards and Technology, Gaithersburg MD, 20899, https://doi.org/10.18434/T4D303, (retrieved September 2, 2022).

Lie, Knut-Andreas. An introduction to reservoir simulation using MATLAB/GNU Octave: User guide for the MATLAB Reservoir Simulation Toolbox (MRST). Cambridge University Press, 2019

Nordbotten, J.M., Benali, B., Both, J.W., Brattekås, B., Storvik, E., Fernø, M.A.; Two-scale image processing for porous media, TiPM SI (2023), submitted

Nordbotten JM, Fernø MA, Flemisch B, Juanes R, Jørgensen M (2022). Final Benchmark Description: FluidFlower International Benchmark Study. Zenodo. https://doi.org/10.5281/zenodo.6807102

Saló-Salgado, L., Haugen, M., Eikehaug, E., Fernø, M.A., Nordbotten J.M., Juanes, R. Direct Comparison of Numerical Simulations and Experiments of CO2 Injection and Migration in Geologic Media: Value of Local Data and Predictability, TiPM SI (2023), submitted

Wentworth, C.K., A Scale of Grade and Class Terms for Clastic Sediments. The Journal of Geology, 1922. 30(5): p. 377-392.

**Supplementary information:**

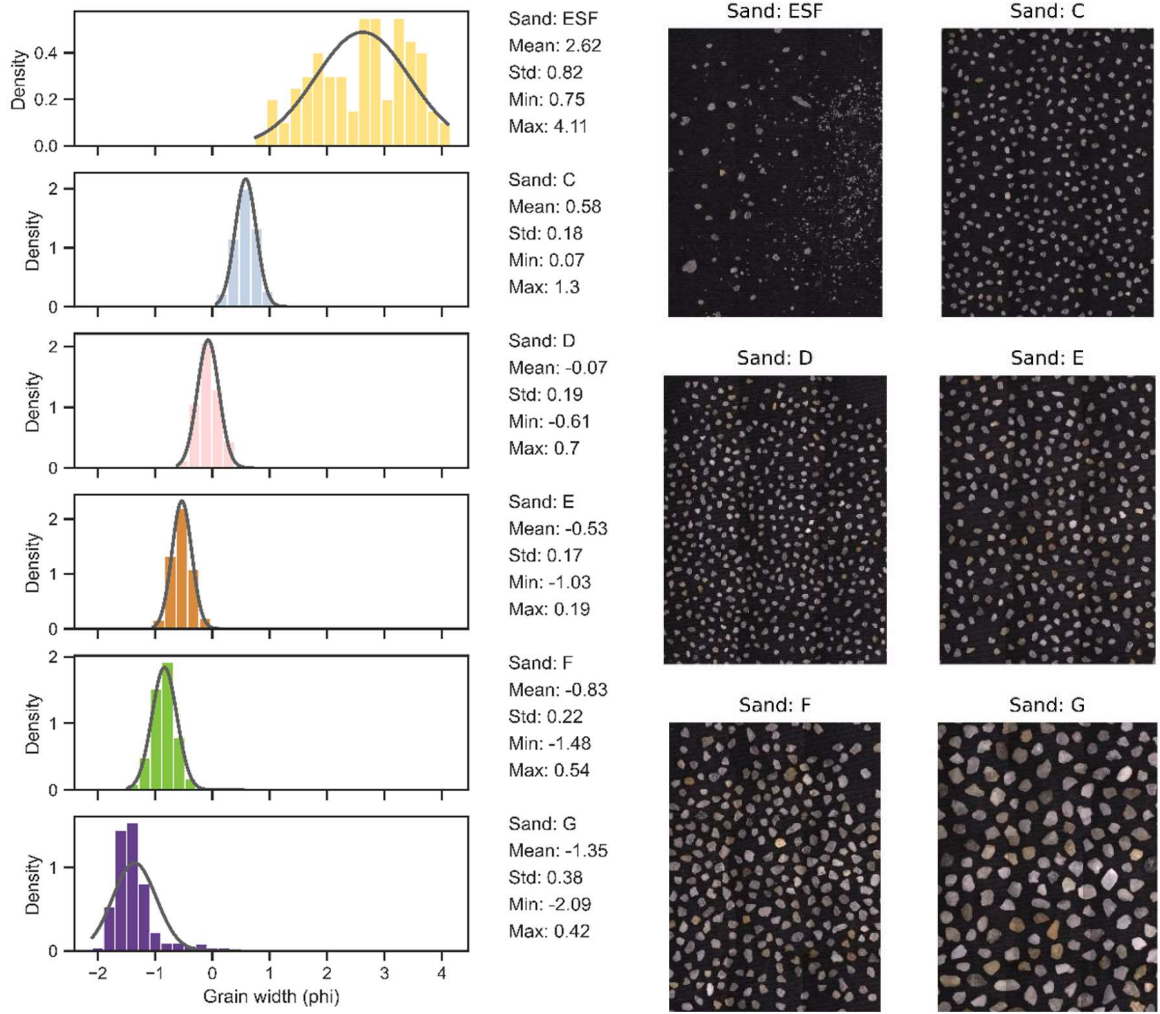

*SI.fig. 1: Left side: Histogram (0.2 binwidth) of grain width distribution for sand used in this study. Statistical information in phi units. Right side: Example of microscope image of each sand used in the analysis.*

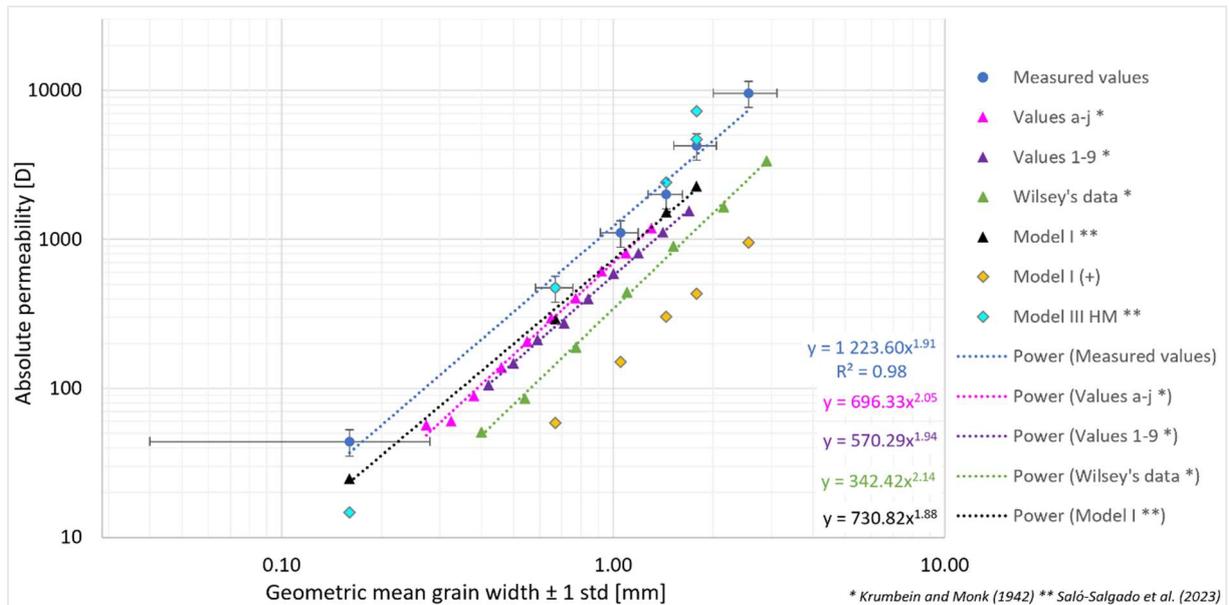

*SI.fig. 2: Measured absolute permeability (blue circles, with ± 20 % uncertainty) plotted against geometric mean of grain width. Five literature dataset are included for comparison: "Values a-j" and "Values 1-9" (porosity = 0.40) and "Wilsey's data" with porosity = 0.40-0.42 (Krumbein and Monk, 1942); and two dataset from **SI.table 2**: "Model I" used as initial values during history match (HM), and "Model III HM" are the final values used to HM experiment AC02. "Model I (+)" is generated from fitting the mathematical model used to obtain input values for Model I to the measured values of grain size and porosity (**Table 2**), and further used it to calculate absolute permeability. This comparison shows that even though the calculated values based on grain size and porosity are much lower, the values needed to HM experiment AC02 are similar to measured values for sand C, E and F.*

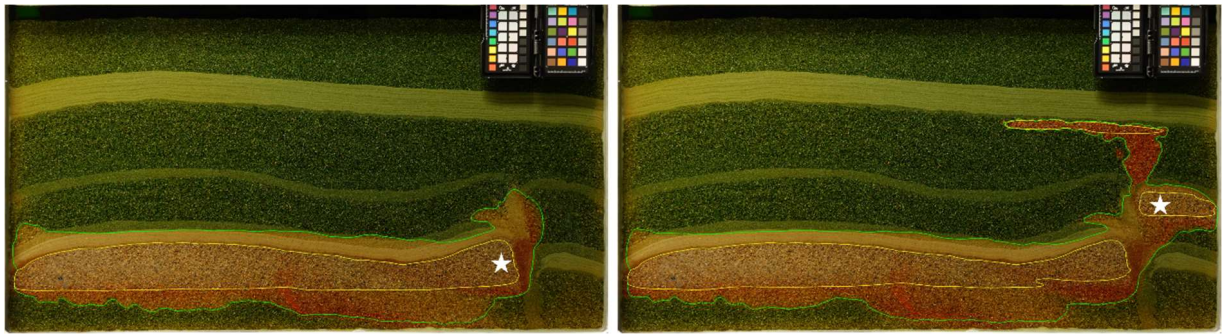

*SI.fig. 3: Capillary entry pressure in horizontal versus vertical geological features in experiment AC07. The sealing capacity for different layers were evaluated by recording under each layer in. The maximum gas column height in the vertical orientation of C-sand in the fault zone co (star symbol in left image) is equivalent to a capillary entry pressure of 6.8 mbar. In the horizontal orientation of C-sand (star symbol in right image), a gas-column height equivalent of 3.1 mbar was recorded. The horizontal C-sand is comparable to measured value in Section 1 (**Table 3**), whereas the vertically deposited C-sand in the fault zone has a higher value, possibly related to the packing and orientation of the grains.*

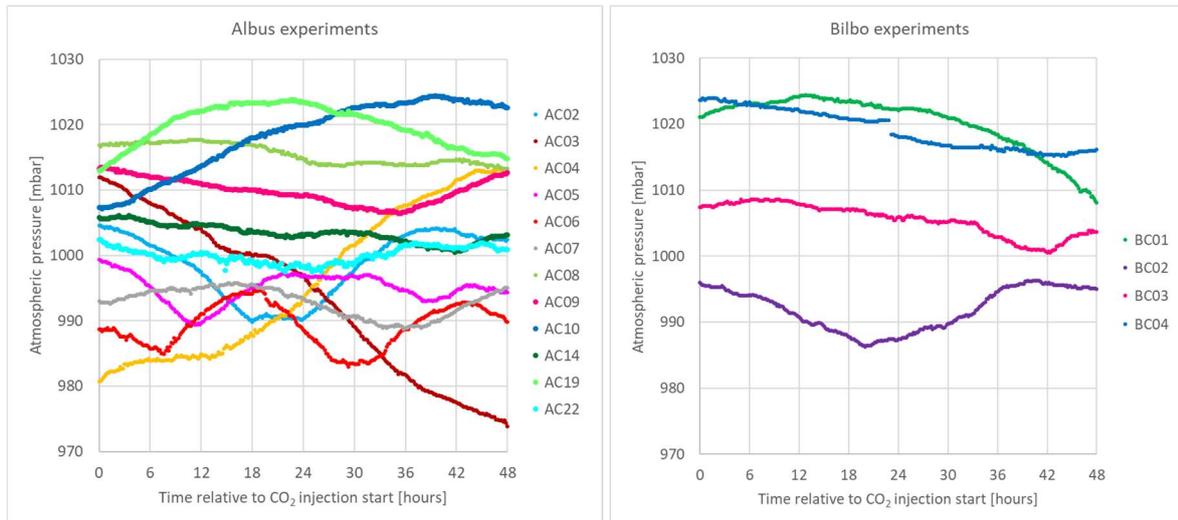

*SI.fig. 4: Variations in atmospheric pressure in the lab during the experiments (Geophysical institute, 2022) from October 2021 to November 2022.*

*SI.table 1: Details about duration [seconds] for each $CO_2$ injection rates used in the experiments. Note that in experiment AC02-AC09, the injection was not scripted and hence there are some differences in injection length at each step.*

| MFC inj. rate [%] | Time [seconds] at each MFC injection rate step | | | | | | | | | | | | | | | |
|---|---|---|---|---|---|---|---|---|---|---|---|---|---|---|---|---|
| | AC02 | AC03 | AC04 | AC05 | AC06 | AC07 | AC08 | AC09 | AC10 | AC14 | AC19 | AC22 | BC01 | BC02 | BC03 | BC04 |
| 0 * | 500 | 503 | 440 | 500 | 500 | 500 | 500 | 500 | 500 | 0 | 0 | 0 | 0 | 0 | 0 | 0 |
| 1 [a] | 60 | 60 | 60 | 60 | 60 | 60 | 60 | 60 | 60 | 60 | 60 | 60 | - | - | - | - |
| 2 | 60 | 60 | 60 | 60 | 60 | 60 | 60 | 60 | 60 | 60 | 60 | 60 | - | - | - | - |
| 5 | 60 | 60 | 60 | 60 | 60 | 60 | 60 | 60 | 60 | 60 | 60 | 60 | - | - | - | - |
| 10 | 60 | 60 | 60 | 60 | 60 | 60 | 60 | 60 | 60 | 60 | 60 | 60 | 60 | 60 | 60 | 60 |
| 15 | 60 | 60 | 60 | 60 | 60 | 60 | 60 | 60 | 60 | 60 | 60 | 60 | 60 | 60 | 60 | 60 |
| 20 | 2751 | 2696 | 2696 | 2689 | 7374 | 15988 | 2700 | 2698 | 2698 | 2700 | 2700 | 2700 | 3793 | 3793 | 3793 | 3793 |
| 15 | 60 | 60 | 60 | 82 | 28 | 28 | 60 | 60 | 60 | 60 | 60 | 60 | 60 | 60 | 60 | 60 |
| 10 | 60 | 60 | 60 | 60 | 81 | 81 | 60 | 60 | 60 | 60 | 60 | 60 | 60 | 60 | 60 | 60 |
| 5 | 60 | 60 | 60 | 60 | 60 | 60 | 60 | 60 | 60 | 60 | 60 | 60 | 60 | 60 | 60 | 60 |
| 2 | 60 | 60 | 60 | 60 | 60 | 60 | 60 | 60 | 60 | 60 | 60 | 60 | - | - | - | - |
| 1 | 60 | 60 | 60 | 60 | 60 | 60 | 60 | 60 | 60 | 60 | 60 | 60 | - | - | - | - |
| 0 ** | 856 | 628 | 823 | 788 | - | - | 1634 | 833 | 834 | 220 | 220 | 220 | 77 | 77 | 77 | 77 |
| 1 | 60 | 60 | 60 | 60 | - | - | 60 | 60 | 60 | 60 | 60 | 60 | - | - | - | - |
| 2 | 60 | 60 | 60 | 60 | - | - | 60 | 60 | 60 | 60 | 60 | 60 | - | - | - | - |
| 5 | 60 | 60 | 60 | 60 | - | - | 60 | 60 | 60 | 60 | 60 | 60 | - | - | - | - |
| 10 | 60 | 60 | 60 | 60 | - | - | 60 | 60 | 60 | 60 | 60 | 60 | 60 | 60 | 60 | 60 |
| 15 | 60 | 60 | 60 | 60 | - | - | 60 | 60 | 60 | 60 | 60 | 60 | 60 | 60 | 60 | 60 |
| 20 | 4473 | 4488 | 4488 | 4484 | - | - | 4500 | 4497 | 4500 | 4500 | 4500 | 4500 | 4372 | 4372 | 4372 | 4372 |
| 15 | 28 | 28 | 28 | 28 | - | - | 28 | 28 | 28 | 21 | 21 | 21 | 60 | 60 | 60 | 60 |
| 10 | 81 | 81 | 81 | 81 | - | - | 81 | 81 | 81 | 81 | 81 | 81 | 60 | 60 | 60 | 60 |
| 5 | 60 | 60 | 60 | 60 | - | - | 60 | 60 | 60 | 60 | 60 | 60 | 60 | 60 | 60 | 60 |
| 2 | 60 | 60 | 60 | 60 | - | - | 60 | 60 | 60 | 60 | 60 | 60 | - | - | - | - |
| 1 | 60 | 60 | 60 | 60 | - | - | 60 | 60 | 60 | 60 | 60 | 60 | - | - | - | - |
| 0 *** | | | | | | | | | | | | | | | | |

[a] Correspond to time = 0 in the analysis and image sets
* Time [sec] from $CO_2$ enter valve to $CO_2$ ramp-up is initiated
** First injection valve closed and moved to second injection port
*** Second injection port closed, and system left until 48 hours from first $CO_2$ injection ramp up

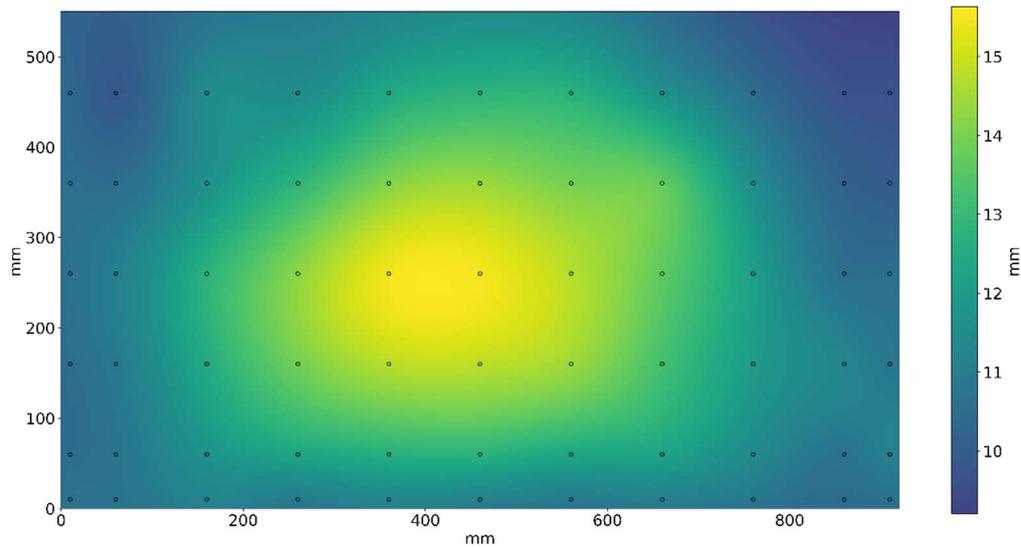

*SI.fig. 5: Depth map for the Albus and Bilbo geometry. Measurements are done on the front side of the Bilbo rig (locations marked with black circles in the image), and assumption is made that the expansion is equal on both sides since Plexiglas is used on both front and back side of the Medium FluidFlower rigs. Because of ongoing experiments in the Albus geometry, it is also assumed that the expansion is the same in both rigs, and the same depth map is therefore used in the Albus geometry but scaled to the correct size. The measurements are performed on the marked circles in the image, and a radial basis interpolation is performed to get the full depth map with values at every pixel, as presented here.*

*SI.table 2: Overview of petrophysical input parameters for simulation Model I to III used to history match (HM) AC02 in the Albus geometry, from Saló-Salgado et al. (2023).*

| Model I | | | | | | | | | | | | | | | |
|---|---|---|---|---|---|---|---|---|---|---|---|---|---|---|---|
| Sand | Porosity | | K [D] | | $S_{wi}$ | | $k_{rel\ gas}$ | | $1-S_g$ | | $k_{rel\ water}$ | | $S_g$ | | $P_c$ | |
| | Initial | HM | Initial | HM | Initial | HM | Initial | HM | Initial | HM | Initial | HM | Initial | HM | Initial | HM |
| ESF | 0.37 | 0.37 | 25 | 6.2 | 0.09 | 0.09 | 0.971 | 0.971 | 0.85 | 0.85 | 0.478 | 0.478 | 1E-5 | 1E-5 | 31.4 | 31.4 |
| C | 0.38 | 0.38 | 293 | 292.8* | 0.03 | 0.03 | 0.971 | 0.971 | 0.84 | 0.84 | 0.478 | 0.478 | 1E-3 | 1E-3 | 9.28 | 4.6 |
| E | 0.39 | 0.39 | 1528 | 1527.8 | 0.01 | 0.01 | 0.971 | 0.971 | 0.84 | 0.84 | 0.478 | 0.478 | 1E-3 | 1E-3 | 4.11 | 0.5 |
| F | 0.39 | 0.39 | 2277 | 2907** | 0.01 | 0.01 | 0.971 | 0.971 | 0.84 | 0.84 | 0.478 | 0.478 | 1E-3 | 1E-3 | 3.37 | 0 |

* in fault 26.9 D
** 2907 D in the middle F layer and 6540 in the lower and top layers

| Model II | | | | | | | | | | | | | | | |
|---|---|---|---|---|---|---|---|---|---|---|---|---|---|---|---|
| Sand | Porosity | | K [D] | | $S_{wi}$ | | $k_{rel\ gas}$ | | $1-S_g$ | | $k_{rel\ water}$ | | $S_g$ | | $P_c$ | |
| | Initial | HM | Initial | HM | Initial | HM | Initial | HM | Initial | HM | Initial | HM | Initial | HM | Initial | HM |
| ESF | 0.44 | 0.44 | 44 | 44 | 0.09 | 0.09 | 0.971 | 0.971 | 0.85 | 0.85 | 0.478 | 0.478 | 1E-5 | 1E-4 | 25.6 | 25.6 |
| C | 0.44 | 0.44 | 473 | 473* | 0.03 | 0.03 | 0.971 | 0.971 | 0.84 | 0.84 | 0.478 | 0.478 | 1E-3 | 1E-3 | 7.81 | 2.6 |
| E | 0.45 | 0.45 | 2005 | 3008 | 0.02 | 0.01 | 0.971 | 0.971 | 0.84 | 0.84 | 0.478 | 0.478 | 1E-3 | 1E-3 | 3.86 | 0.6 |
| F | 0.44 | 0.44 | 4259 | 4259** | 0.01 | 0.01 | 0.971 | 0.971 | 0.84 | 0.84 | 0.478 | 0.478 | 1E-3 | 1E-3 | 2.62 | 0 |

* in fault 158 D
** 4259 D in the middle F layer, and 6814 in the lower and top layers

| Model III | | | | | | | | | | | | | | | |
|---|---|---|---|---|---|---|---|---|---|---|---|---|---|---|---|
| Sand | Porosity | | K [D] | | $S_{wi}$ | | $k_{rel\ gas}$ | | $1-S_g$ | | $k_{rel\ water}$ | | $S_g$ | | $P_c$ | |
| | Initial | HM | Initial | HM | Initial | HM | Initial | HM | Initial | HM | Initial | HM | Initial | HM | Initial | HM |
| ESF | 0.44 | 0.44 | 44 | 15 | 0.32 | 0.32 | 0.971 | 0.971 | 0.86 | 0.86 | 0.38 | 0.38 | 1E-4 | 1E-4 | 15.0 | 30*** |
| C | 0.44 | 0.44 | 473 | 473* | 0.14 | 0.14 | 0.971 | 0.971 | 0.90 | 0.90 | 0.596 | 0.596 | 1E-3 | 1E-3 | 3.0 | 4.5 |
| E | 0.45 | 0.45 | 2005 | 2406 | 0.12 | 0.12 | 0.971 | 0.971 | 0.94 | 0.94 | 0.744 | 0.744 | 1E-3 | 1E-3 | 0.3 | 0.3 |
| F | 0.44 | 0.44 | 4259 | 4685** | 0.12 | 0.12 | 0.971 | 0.971 | 0.87 | 0.87 | 0.512 | 0.512 | 1E-3 | 1E-3 | 0.0 | 0 |

* in fault 118 D
** 4685 D in the middle F layer, and 7240 D in the lower and top layers
*** this values could also have been kept as the measured value with similar result

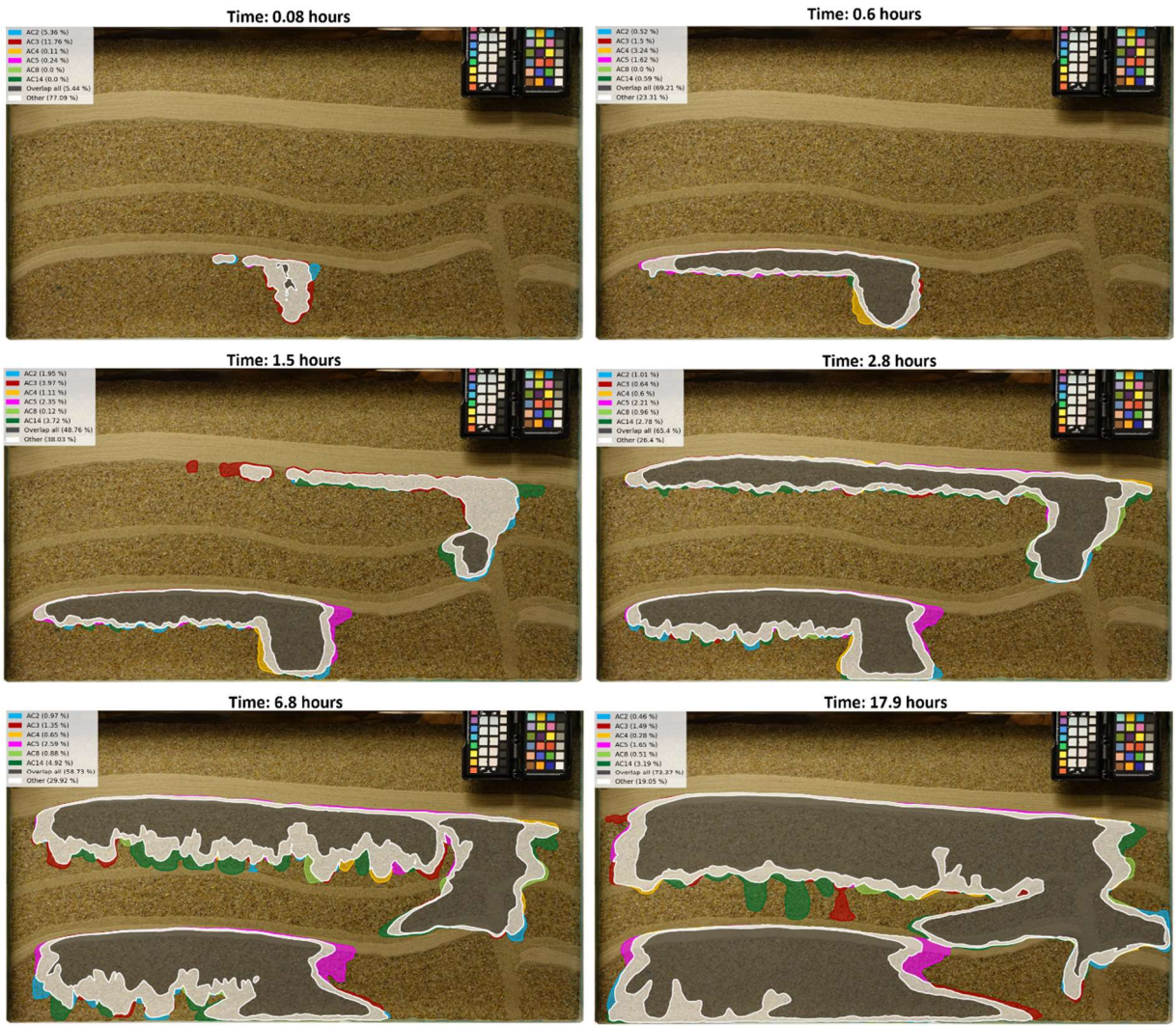

*SI.fig. 6:* Overlapping timeseries for CO$_2$ injection experiments with different degree of degassing the aqueous phase in the Albus geometry

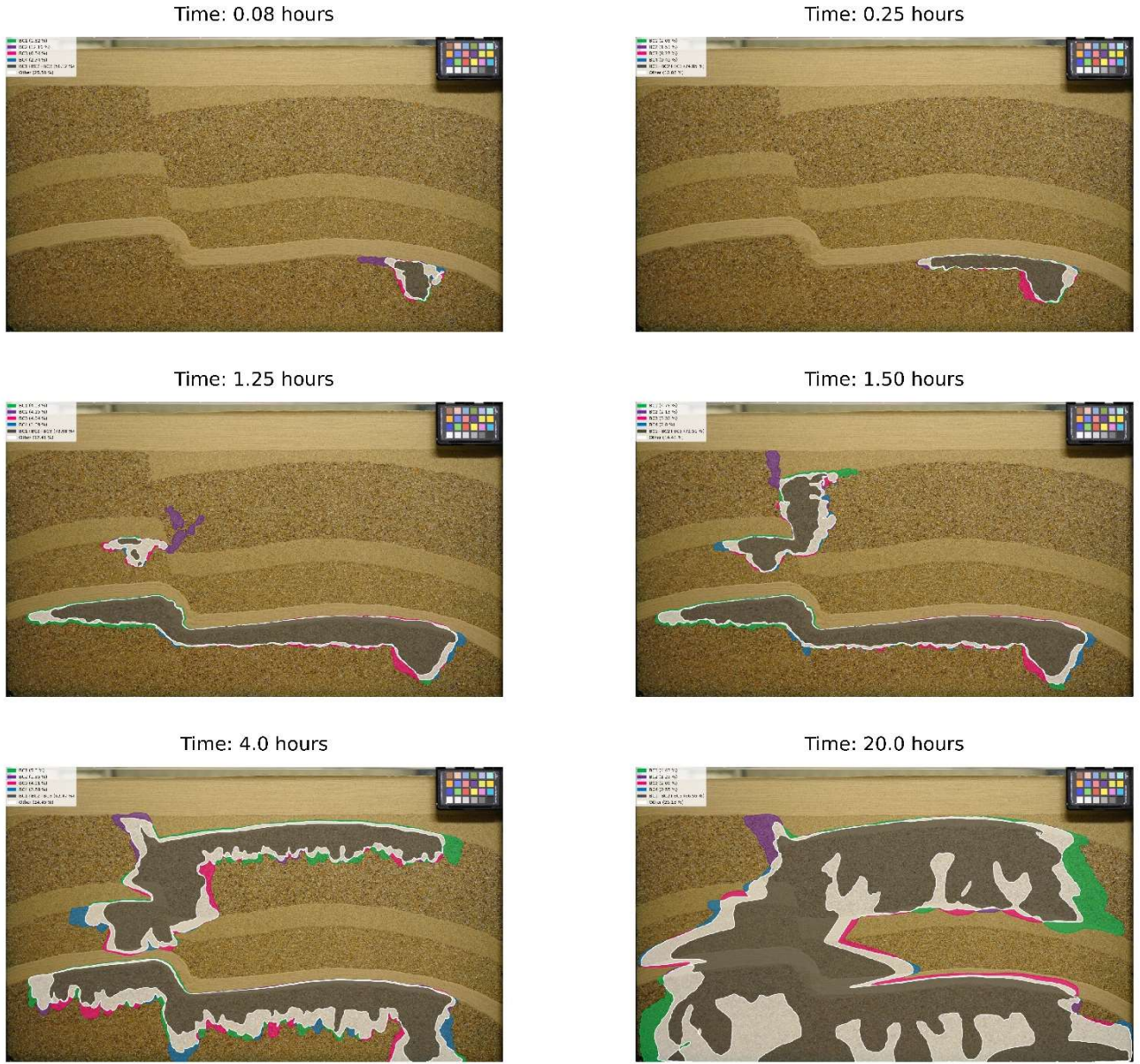

SI.fig. 7: Time series of spatial uniqueness and overlap for experiments BC01-04: BC01 (green), BC02 (purple), BC03 (pink) and BC04 (blue); overlap for BC01-03; other overlap combinations (white). Refer to **SI.table 3** for numbers. The gravitational fingers in the Middle F layer (after 4.0, from the first injection port) propagated longest for BC04 (pH mix 3, high pH) with more dissolved $CO_2$ and less mobile $CO_2$ exceeding the spill-point for the "smeared" part of the fault. This resulted in higher concentration of $CO_2$ in the dissolved fluid and thereby faster propagating fingers. The difference between BC01, BC02 and BC03 relates to methyl red precipitation (in pH mix 1, BC01 and BC02), which precipitates at pH ~6.2 (Eikehaug et al., this issue). The overlap (gray) after 4 and 20 hours is limited by the spatial distribution of experiment BC02 because methyl red precipitates create a low permeability boundary which is preventing gravitational fingers to develop along the whole interface.

*SI.table 3: Numbers from label in SI.fig. 7.*

| Time [h] | BC01 | BC02  | BC03 | BC04 | BC01-03 | Other |
|----------|------|-------|------|------|---------|-------|
| 0.08     | 1.8% | 12.2% | 6.5% | 2.7% | 51.1%   | 25.6% |
| 0.25     | 2.1% | 1.5%  | 8.3% | 0.4% | 74.9%   | 12.9% |
| 1.25     | 4.0% | 4.2%  | 4.0% | 1.7% | 74.7%   | 12.4% |
| 1.5      | 4.8% | 2.1%  | 3.3% | 2.8% | 72.5%   | 14.5% |
| 4.0      | 5.0% | 1.4%  | 4.0% | 2.7% | 62.5%   | 24.5% |
| 20.0     | 4.4% | 1.2%  | 2.1% | 0.6% | 66.6%   | 25.1% |

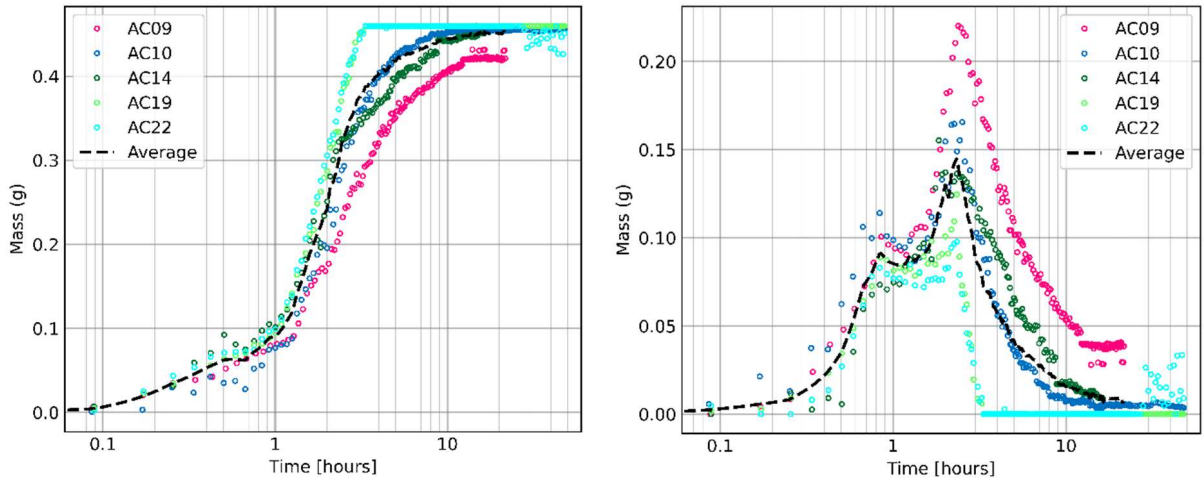

*SI.fig. 8: Selection of $CO_2$ injection experiments in the Albus geometry with higher degree of vacuuming pH-indicator. Note that AC09 uses pH-indicator mix number 2, and as for BC03, a larger amount of residual gas is detected/segmented by DarSIA compared to pH-indicator mix number 1 and 3, see also discussion of **Fig. 13**.*

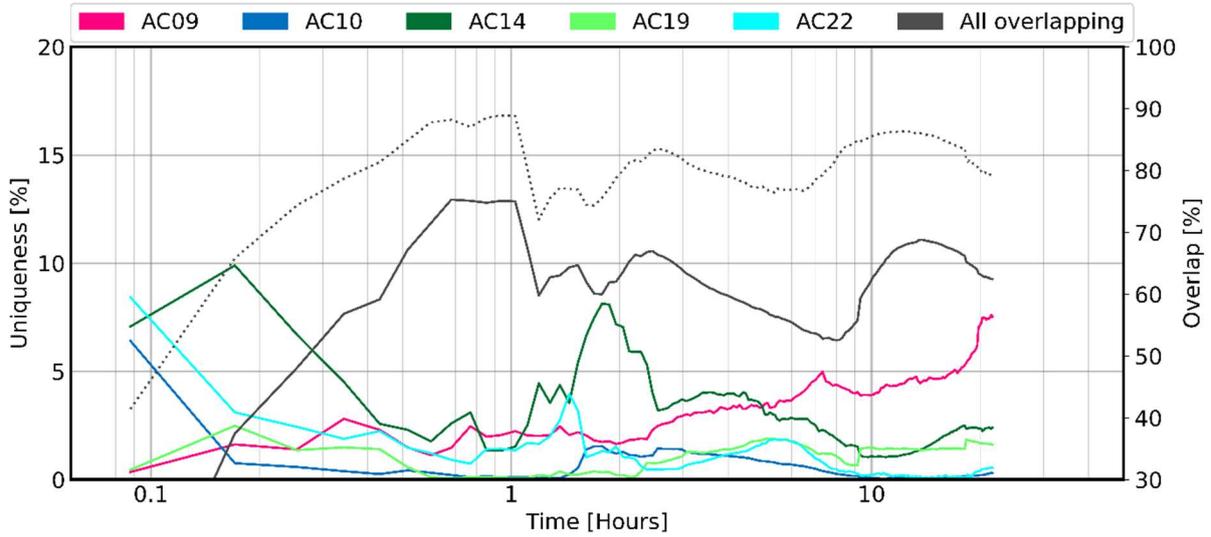

*SI.fig. 9: Uniqueness of $CO_2$ injection experiments in the Albus geometry. The grey line is share of pixels where all experiments are overlapping, and the dotted line is where experiment AC09 and AC14 are overlapping.*

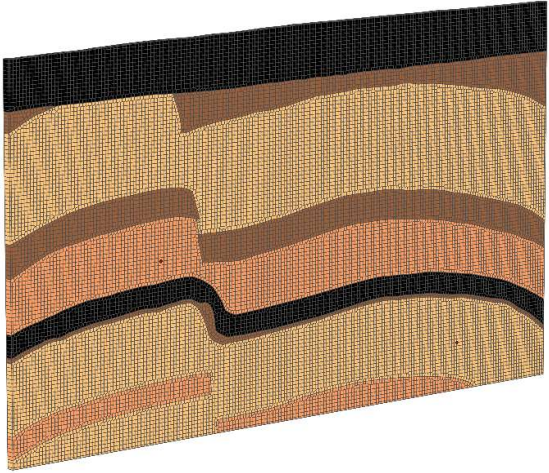

*SI.fig. 10: 3D grid used in MRST simulations of the Bilbo geometry. Dimensions (depth, width, height) are: 10.5 x 934 x 533 mm. The grid has a cell size of 5mm and a total of 20,470 cells. Note that a constant depth of 10.5 mm is used.*

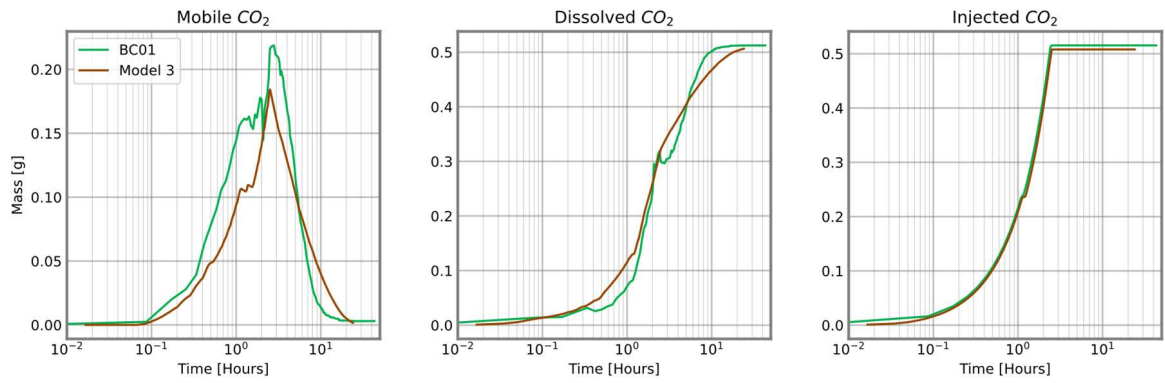

*SI.fig. 11: Comparison of calculated mass of mobile CO2, dissolved CO2 and injected CO2 in Model III and experiment BC01.*

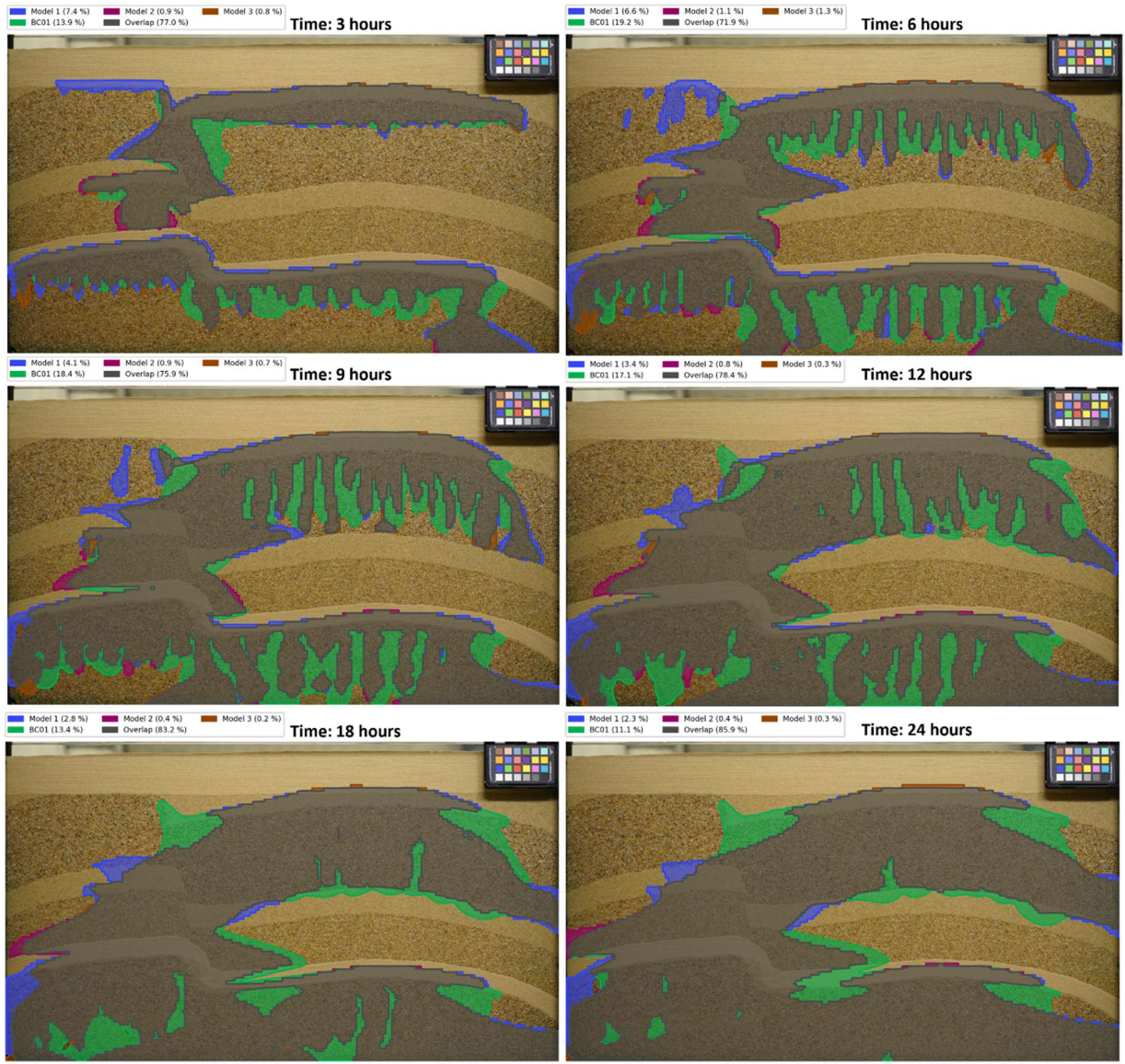

SI.fig. 12: Time series of spatial uniqueness and overlap for Model I (blue), Model II (dark pink), Model III (brown) and experiment BC01 (green); overlap (grey) where minimum two of them overlap each other. Each color represents the spatial distribution of mobile and dissolved $CO_2$. Contours of the simulation results are obtained from gas concentration maps with a threshold value of 0.21 kg/m3 (15% of the maximum value around 1.4 kg/m3).